\documentclass[letterpaper,11pt]{article}
\usepackage{lmodern}
\usepackage{graphicx}
\usepackage{caption}
\usepackage{verbatim}
\usepackage{multicol}
\usepackage{amsmath}
\usepackage{xcolor}
\usepackage{hyperref}
\usepackage[capitalise]{cleveref}
\hypersetup{pdfborder = {0 0 0}}
\hypersetup{pdfborder = {0 0 0}, colorlinks=true, linkcolor=blue!60!black, citecolor=blue!60!black, urlcolor=black}
\usepackage{csquotes}
\usepackage{subcaption}
\usepackage{booktabs}
\usepackage{tabularx}
\usepackage{multirow}
\usepackage{amsfonts}
\usepackage{bbm}
\usepackage{geometry}
\usepackage{enumitem}
\usepackage{wesu}
\usepackage{empheq}
\usepackage{natbib}
\usepackage{mathtools}
\setcitestyle{authoryear,open={(},close={)}} 
\usepackage{minibox}

\definecolor{ABlue}{rgb}{0.0,0.7,0.8}

\title{\textbf{On Coordinate Frames in Axisymmetric Static Vacuum Spacetimes and Implications for Observations}}
\date{February 11, 2025}

\author{A. Seifert}

\newcommand{\diff}{\mathrm{d}}
\newcommand{\efe}{Einstein field equations}
\newcolumntype{C}{>{\centering\arraybackslash}X}
\newcolumntype{R}{>{\raggedleft\arraybackslash}X}
\newcolumntype{L}{>{\raggedright\arraybackslash}X}
\newcommand{\appropto}{\mathrel{\vcenter{
  \offinterlineskip\halign{\hfil$##$\cr
    \propto\cr\noalign{\kern1pt}\sim\cr\noalign{\kern-1pt}}}}}
\newcommand{\breakline}{ \nonumber\\ &\ \ \ \ \ \ \ \ }

\newcommand{\rschd}{\mathbbm{r}}
\newcommand{\risoschd}{r}
\newcommand{\rlp}{\rho}
\newcommand{\zlp}{\zeta} 
\newcommand{\rma}{p}
\newcommand{\zma}{z}
\newcommand{\ran}{\mathfrak{r}}
\newcommand{\zan}{\mathfrak{z}}

\begin{document}
\maketitle

\noindent
\begin{minipage}[b]{\textwidth}
    \noindent
    \begin{abstract}
        While a physical theory should be independent of the coordinate frame chosen by any observer, the observations themselves in fact depend on the choice of coordinates. In particular, different coordinate frames reflect different symmetries seen by a local observer. 
        In this work, we discuss the applicability of different coordinate choices and the resulting line elements for static axisymmetric vacuum spacetimes.
        We find that the effective potential experienced by a local observer in the low-velocity limit is highly dependent on the form of the line element and thus on the coordinates chosen in the description. For example, this affects the form of a rotation curve expected by such an observer. We thus conclude that it is important to review the choices of local (coordinate frame of the observer) and global symmetries carefully to understand observations from a generally relativistic point of view.

    \end{abstract}
\end{minipage}

\section{Introduction}

General relativity (GR) has proven to be a very successful theory of gravity ever since its introduction \citep{Einstein1915SPAW.......844E}, explaining and predicting phenomena such as Mercury's perihelion shift \citep{clemence_mercury_1947RvMP...19..361C}, black holes \citep{Schwarzschild1916SPAW.......189S,Kerr:1963PhRvL..11..237K,gravity_blackholeobs_2018A&A...615L..15G}, gravitational waves \citep{Abbott_gravwaves_2016PhRvL.116f1102A} and providing the foundation of the standard model of cosmology.

To derive the geometry of a given spacetime manifold from the \efe, it is crucial to impose symmetries on the ansatz for the metric. Examples for such spacetime manifolds are the Schwarzschild solution \citep{Schwarzschild1916SPAW.......189S} which is spherically symmetric and static, and the Kerr metric \citep{Kerr:1963PhRvL..11..237K} which is axisymmetric and stationary. Both of these solutions are vacuum solutions, i.e. they describe spacetimes without any mass content in the domain of definition. Although the Schwarzschild solution is built on the unphysical assumptions to have a static vacuum spacetime, it reduces to Newtonian gravity in the low-velocity and weak-field limit. 

Situations where this limit is applicable, such as in galaxies, are generally assumed to be described by Newtonian gravity. When only considering baryonic matter, the Newtonian prediction fails to explain the rotation curves observed in galaxies \citep[for a review see][]{Bertone_historyDM:2018RvMP...90d5002B}. This contradiction is commonly resolved by considering a halo of dark matter that contributes to the dynamics in the galaxy by gravitational interaction only. However, particles constituting this dark matter halo have not been found to date and the nature of dark matter remains an open question in modern physics \citep[for a review see][]{Bertone_searchDM:2018Natur.562...51B}. Other approaches to explain the ``missing mass'' are the theories of modified Newtonian dynamics \citep[MOND, see e.g.,][]{Milgrom:1983ApJ...270..365M,McGaugh2016PhRvL.117t1101M} and contributions of general relativistic self-interaction effects \citep[GR-SI, see e.g.,][]{Deur:2009ya,deur_2020}. While the latter does not rely on a specific metric, multiple attempts towards a full relativistic treatment of galaxies have been made in the past years \citep[for a review see][]{Re_Galoppo:2024arXiv240303227R}, both from a analytical perspective and by comparison to observations \citep[e.g.,][]{cooperstock:2005astro.ph..7619C,crosta:2020MNRAS.496.2107C}. Lately, \citet{Beordo:Crosta:2024MNRAS.529.4681B} investigated the applicability of a stationary and axisymmetric class of general relativistic solutions to the Milky Way and found that general relativistic effects can contribute substantially to the rotation curve. However, their class of solutions is based on a stationary metric allowing for frame-dragging effects \citep[see e.g.,][]{Astesiano2022,Stephani_Kramer_MacCallum_Hoenselaers_Herlt_2003} which require further investigation. 

In this work, we consider the effect of symmetries on the motion of particles in the frame of a local observer. To this end, we consider a static vacuum metric and analyse its low-velocity limit. This is completely analogous to how the Newtonian approximation emerges from the static spherically symmetric vacuum solution given by the Schwarzschild metric, except for the different symmetry conditions.

\section{Applicability of Coordinate Frames}

The theory of general relativity and the \efe\ can be expressed in coordinate-free form 
\begin{align}
    \mathbf{G} = \frac{8\pi G}{c^4} \mathbf{T},\label{eq:efe-tensor}
\end{align}
where $\mathbf{G} = \mathbf{R} - \frac{\mathrm{tr} \mathbf{R}}{2} \mathbf{g}$ is the Einstein tensor, $\mathbf{g}$ is the metric tensor, $\mathbf{T}$ is the energy-momentum tensor, and $G$ is the gravitational constant. While the physical theory should not depend on any choice of coordinates, the observations made will depend on the observer's coordinate frame. 

\subsection{Examples from the Schwarzschild Spacetime}

As an example, consider the Schwarzschild solution, which is the unique spherically symmetric solution to the vacuum \efe\ \citep{Schwarzschild1916SPAW.......189S,Birkhoff1923rmp..book.....B}. The most common choice of coordinates for this solution are the Schwarzschild coordinates,
\begin{align}
    \diff s^2 &= -(1 - 2\Psi) c^2 \diff t^2 + \frac{\diff \rschd}{1 - 2 \Psi} + \rschd^2 (\diff \vartheta^2 + \sin^2 \vartheta \diff \varphi^2),\label{eq:schd-standard}\\
    \Psi &= \frac{MG}{c^2} \frac{1}{\rschd}.
\end{align}
In this frame, two spacetime singularities are present, namely $\rschd = 0$ and $\rschd = \frac{MG}{c^2}$. However, the second of these singularities, corresponding to the event horizon at the Schwarzschild radius $R_S = \frac{MG}{c^2}$, is a coordinates singularity and not a curvature singularity (cf. \cref{app:kretschmann}). Thus, this coordinate frame is not applicable close to the Schwarzschild horizon but only to more distant observers.

Instead, an observer falling into a Schwarzschild black hole may consider the metric in Painlevé-Gullstrand coordinates \citep{Painleve1921CR....173..677P,Gullstrand:1922tfa},
\begin{align}
    \diff s^2 = -(1 - 2\Psi) c^2 \diff T^2 + 2\sqrt{2 \Psi}\diff T\diff \rschd + \diff \rschd^2 + \rschd^2 (\diff \vartheta^2 + \sin^2 \vartheta \diff \varphi^2).
\end{align}
In these coordinates, the observer crosses the event horizon at $\rschd = R_S$ with velocity $v = \frac{\diff \rschd}{\diff T} = \beta c = \sqrt{2\Phi}c$, whereas for a distant observer the velocity $v= \frac{\diff \rschd}{\diff t}$ tends to zero when approaching the horizon. The Painlevé-Gullstrand coordinates are not singular at $R = R_S$ and can thus be used to describe observers close to the horizon.

Neither the Schwarzschild coordinates nor the Painlevé-Gullstrand coordinates can describe the full space-time manifold with a single chart of coordinates. However, this is possible in the frame of the Kruskal-Szekeres coordinates \citep[e.g.,][]{misnerthornewheeler1977gvi..book.....M},
\begin{align}
    \diff s^2 &= 32 \Psi^3 \rschd^2 e^{-\frac{1}{2\Psi}} (-\diff \eta^2 + \diff \chi^2) + \rschd^2 (\diff \vartheta^2 + \sin^2 \vartheta \diff \varphi^2),\label{eq:kruskal}\\
    \eta^2 - \chi^2 &= \left(1 - \frac{1}{2\Psi}\right) e^{\frac{1}{2\Psi}},
\end{align}
the latter giving an implicit definition of $\rschd = \rschd(\eta, \chi)$ in the new coordinate frame. Note that in this choice of coordinates, the components of the metric depend on both the time-like coordinate $\eta$ and a space-like coordinate $\chi$, it is thus not the preferred choice of a stationary observer with a corresponding time-like Killing vector.

Going back to coordinate frames relevant to an observer, another obvious choice would be an isotropic coordinate frame, as this is what will be experienced to a distant observer. This is found for the line element
\begin{align}
    \diff s^2 &= -\frac{\left(1 - \frac{\Phi}{2}\right)^2}{\left(1 + \frac{\Phi}{2}\right)^2} c^2 \diff t^2 + \left(1 + \frac{\Phi}{2}\right)^4 \left(\diff \risoschd^2 + \risoschd^2 \diff \vartheta^2 + \risoschd^2 \sin^2 \vartheta \diff \varphi^2\right)\label{eq:schd-isotr-metric},\\
    \Phi &= \frac{MG}{c^2} \frac{1}{\risoschd}.
\end{align}
Note that the potential $\Phi$ is different from $\Psi$ as it relies on the isotropic radius $\risoschd$ instead of $\rschd$ from the Schwarzschild coordinates in \cref{eq:schd-standard}. Thus, this line element is singular in $\risoschd = R_S$ and $\risoschd = 0$. It is particularly important in the Newtonian limit, as Newtonian observers will expect an isotropic local coordinate frame. 

Notably, for small $\Phi$ the line element in \cref{eq:schd-isotr-metric} reduces to the isotropic Newtonian metric
\begin{align}
    \diff s^2 &= -\left(1 - 2\Phi\right) c^2 \diff t^2 + \left(1 + 2 \Phi\right) \left(\diff \risoschd^2 + \risoschd^2 \diff \vartheta^2 + \risoschd^2 \sin^2 \vartheta \diff \varphi^2\right).\label{eq:newt-isotr-metr}
\end{align}
It has to be emphazised that unlike the line elements presented before, this is \textit{not} a exact solution to the vacuum \efe, but only solves them approximately in the Newtonian limit, i.e. $\Phi \ll 1$.

\subsection{Choosing Suitable Coordinate Frames}\label{sec:coordchoice-schd}

The coordinate frame suitable for the description of a physical situation is highly dependent on the observers' assumptions. Such assumptions include limiting behaviours as well as symmetry conditions. In particular, a solution explaining the spacetime of an isolated physical object (e.g. a black hole) is expected to approach the flat metric for the limiting case of large distances from the object. This is the case for the Schwarzschild metric as the line elements presented above approach the Minkowski line element for large radii. 

Furthermore, \cref{eq:schd-standard,eq:schd-isotr-metric} also reflect the radial symmetry. However, these two metrics differ in the local situation experienced by the observer: In the Schwarzschild coordinates, \cref{eq:schd-standard}, the relative scaling of the radial and angular coordinates varies with the radius. In contrast to this, the relative scaling in the isotropic coordinates, \cref{eq:schd-isotr-metric} is the same in every spatial direction. As this is the situation expected for a Newtonian observer, it is the coordinate frame preferred to take the Newtonian limit. Note that theoretically the Newtonian limit could also be taken starting from the Schwarzschild coordinates, \cref{eq:schd-standard}, but in this frame the isotropy assumption is no fulfilled, as the spatial metric depends on the radial coordinate in a non-conformal way.

Apart from the different local appearance of the coordinate frames, their domain of definition differs as well. This can be seen clearly from the singularities present in the different coordinate frames. In particular, the line element in \cref{eq:schd-standard} possesses a coordinate singularity in $\rschd = R_S$ and this thus not suitable for a description there. In contrast, the Kruskal-Szekeres coordinates, \cref{eq:kruskal}, show no singularities and are applicable to the full manifold, but however the symmetries of the spacetime are not reflected in this choice of coordinates.

However, it is in general not a problem if the coordinate frame does not cover the full spacetime manifold. The \efe\ as given in \cref{eq:efe-tensor} are differential equations in the metric tensor $\mathbf{g}$, thus, as derivatives have to be taken locally, \emph{a priori} the solutions are constructed locally as well.

To conclude, a suitable coordinate frame has to be chosen to accomodate for the need of the observer in question in terms of symmetries. Although the coordinate frames are constructed locally, the limiting behaviour has to be taken into account when considering more general observers, as the symmetries of the system are global. With this in mind, we want turn to study axisymmetric metrics.

\section{Axisymmetric Metrics}\label{sec:axisym}

A general axisymmetric line element can be written as 
\begin{align}
    \diff s^2 &= - e^{2a(\ran, \zan)} c^2 \diff t^2 + e^{2 b(\ran, \zan)} \diff \ran^2 + e^{2f(\ran, \zan)} \ran^2 \diff \varphi^2 + e^{2 h(\ran, \zan)} \diff \zan^2 \label{eq:2ansatz}.
\end{align}
where the coordinates $t$ and $\varphi$ can be defined from the Killing vectors $\partial_t, \partial_\varphi$, respectively. If $\partial_\zan$ is a Killing vector as well, we can use it to define the $\zan$ coordinate. This is convenient, as the functions $a, b, f, h$ will not depend on this coordinate if $\partial_\zan$ is a Killing vector. 
However, this does not fix the full coordinate frame, but different radial coordinates can be chosen to accomodate for different choices of the coefficients. In this work, we want to study a particular choice of metric in different coordinate frames.

This choice of metric is a vacuum solution with a $\partial_\zan$ Killing vector. It can be derived in two different ways. On the one side, its symmetries can be expressed in terms of three Killing vectors, $\partial_t, \partial_\varphi, \partial_\zan$. Furthermore, we consider an observer, for whom the local coordinate frame appears to be of cylindrical form, i.e. the coefficients of $\diff \ran$ and $\ran \diff \varphi$ agree, $b(\ran, \zan) = f(\ran, \zan)$. One can then compute the Einstein tensor in terms of the functions $a, b, f, h$ as done in \cref{app:efe}. According to \cref{app:unique}, the vacuum \efe\ have a unique solution in this case, which we will call the \textit{cylinder solution}.

\subsection{Exact Static Solutions with $\partial_\zan$ Killing vector}\label{sec:cyl-vs-lp}

In the literature, \citep[see e.g.,][and Eq.~20.3 of \citeauthor{Stephani_Kramer_MacCallum_Hoenselaers_Herlt_2003}, \citeyear{Stephani_Kramer_MacCallum_Hoenselaers_Herlt_2003}]{Ernst1968PhRv..167.1175E,quevedo1990ForPh..38..733Q,Galoppo2023arXiv231015382G}, the most common form of an axisymmetric metric is
\begin{align}
    \diff s^2 = e^{-2U}\left(e^{2k}(\diff \rlp^2 + \diff \zlp^2) + \rlp^2 \diff \varphi^2\right) - e^{2U}(c\diff t + A\diff \varphi)^2\label{eq:lewis-papapetrou},
\end{align}
which defines the Lewis-Papapetrou form. Choosing $A = 0$, i.e. a static solution, this has the form of \cref{eq:2ansatz} for a specific coordinate system with radial and axial coordinate denoted by $(\ran, \zan) = (\rlp, \zlp)$. Defining the Ernst potential $V = e^{2U} = e^{2a(\rlp, \zlp)}$, this must satisfy the Ernst equation \citep{Ernst1968PhRv..167.1175E,quevedo1990ForPh..38..733Q}
\begin{align}
    V\left(V_{,\rlp\rlp} + V_{,\zlp\zlp} + \frac{V_{,\rlp}}{\rlp}\right) = {V_{,\rlp}}^2 + {V_{,\zlp}}^2,\label{eq:ernst}
\end{align}
in order to fulfill the \efe. In general, $U$, $k$ can depend on both coordinates $\rlp$ and $\zlp$ ($\partial_t, \partial_\varphi$ are Killing vectors). However, in this section we want to concentrate on static metrics, i.e. $A=0$, with an additional Killing vector $\partial_\zeta$.
One such solution is given by 
\begin{align}
    \diff s^2 &= - \frac{\rlp^4}{B^2} c^2\diff t^2 + \frac{B^2}{\rlp^4} \left(\frac{16 \rlp^8}{B^4 C^2}(\diff \rlp^2 + \diff \zlp^2) + \rlp^2 \diff \varphi^2\right)\label{eq:cyl-lp}.
\end{align}
In this choice of coordinates, we have
\begin{align}
    e^{2b(\rlp)} = e^{2h(\rlp)} &= \frac{B^2}{\rlp^4}\frac{16 \rlp^8}{B^4 C^2} = \frac{16 \rho^4}{B^2 C^2},\\
    e^{2b(\rlp)} \neq e^{2f(\rlp)} &= \frac{B^2}{\rlp^4},
\end{align}
i.e. the coefficients of $\diff \rlp^2$ and $\diff \zlp^2$ agree but those of $\diff \rlp^2$ and $\rho^2 \diff \varphi^2$ do not.
However, we can transform this line element into a coordinate frame where $e^{2b(\rma)} = e^{2f(\rma)}$. To this end, we choose the new coordinates $(\rma, \zma)$ such that
\begin{align}
    \frac{\rlp^4}{B^2} &= C \ln \frac{\rma}{R}, & \zlp &= \frac{C\sqrt{E}}{4} \zma,\label{eq:coordchange}\\
    \diff \rlp &= \frac{B^2 C}{4 \rma \rlp^3} \diff \rma, &
    \diff \zlp &= \frac{C\sqrt{E}}{4} \diff \zma,
\end{align}
for constants $B, C, E, R$. Note that $\zma$ differs from the coordinate $\zlp$ only by a constant scale factor, thus $\partial_\zma$ is a Killing vector.
The resulting line element reads
\begin{align}
    \diff s^2 &= - \left(C\ln \frac{\rma}{R}\right) c^2 \diff t^2 + \frac{B}{\rma^2\sqrt{C \ln \frac{\rma}{R}}} \left(\diff \rma^2 + \rma^2 \diff \varphi^2\right) + E\left(C\ln \frac{\rma}{R}\right) \diff \zma^2. \label{eq:result-withoutz-R}
\end{align}
In these coordinates, we have
\begin{align}
    e^{2b(\rma)} \neq e^{2h(\rma)} &= E\left(C\ln \frac{\rma}{R}\right),\\
    e^{2b(\rma)} = e^{2f(\rma)} &= \frac{B}{\rma^2\sqrt{C \ln \frac{\rma}{R}}},
\end{align}
The line element \cref{eq:result-withoutz-R} fulfills the vacuum \efe\ as it is obtained from \cref{eq:cyl-lp} by coordinate transformation. This can also be shown explicitly by considering the \efe\ for the ansatz \cref{eq:2ansatz} derived in \cref{app:efe}. 

\subsection{Comparison of Coordinate Frames}\label{sec:coordchoice}

\begin{figure}[h]
    \centering
    \includegraphics[width=0.43\textwidth]{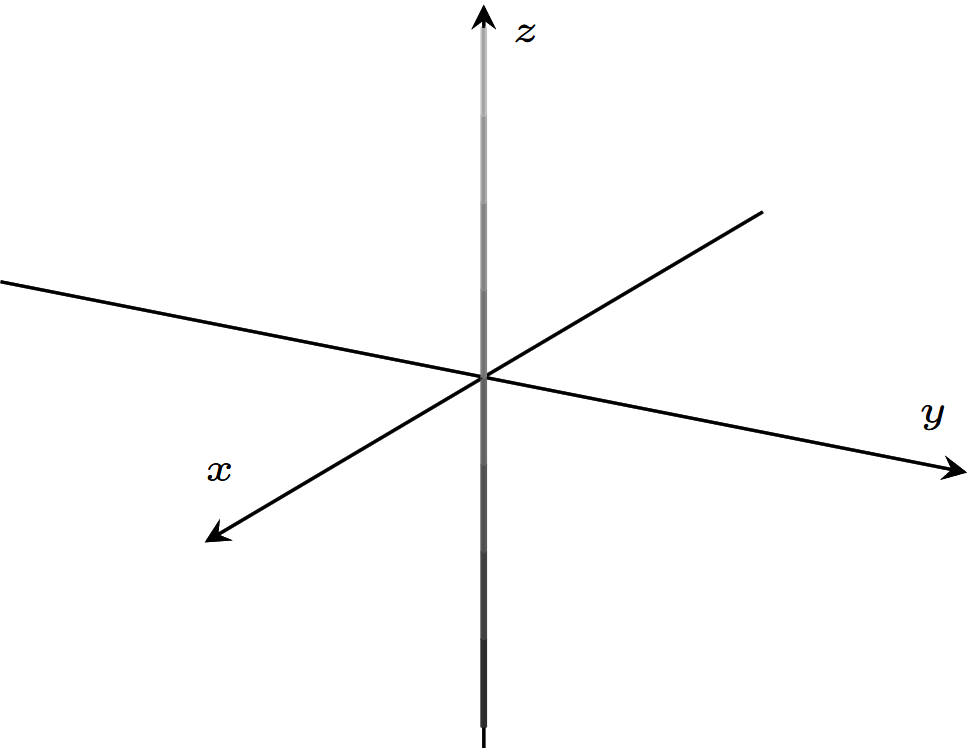}
    \includegraphics[width=0.45\textwidth]{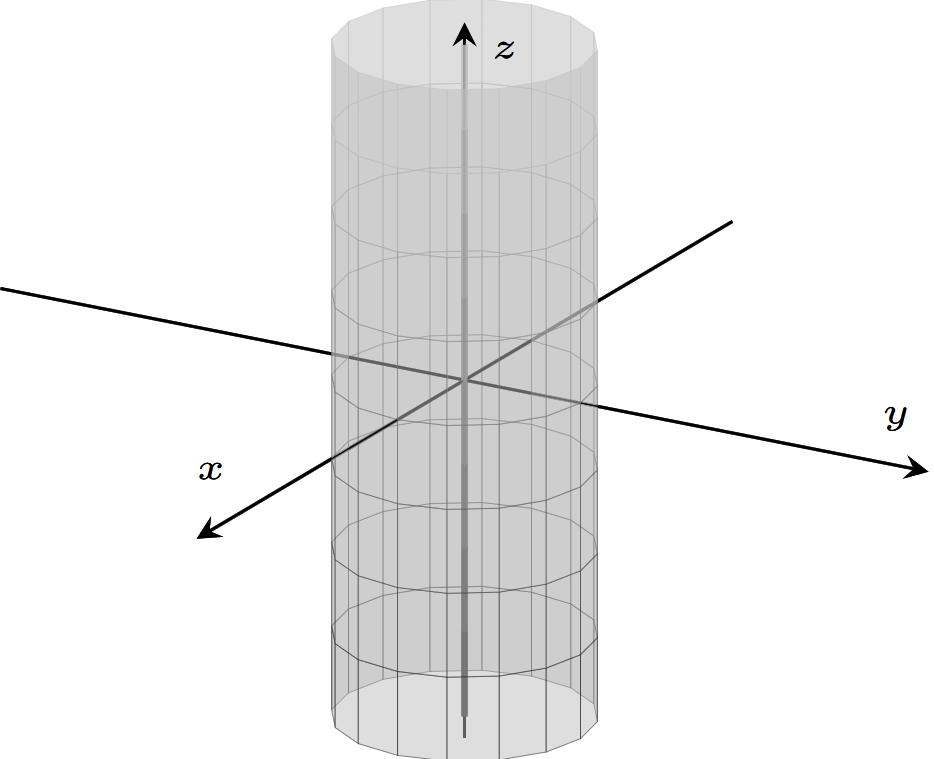}
    \caption{Singularities in the settings described by the line elements in \cref{eq:cyl-lp} (left) and \cref{eq:result-withoutz-R} (right). Due to the appearance in this plot, the coordinates chosen for the right are called the \textit{cylinder frame}.}
    \label{fig:singularities}
\end{figure}

The line elements presented in \cref{sec:cyl-vs-lp} differ in two main properties, the symmetries reflected by the choice of coordinates, and the description of the singularities. In terms of the singularities the line element from \cref{eq:cyl-lp} is singular in $\rlp = 0$ only, which is a curvature singularity (cf. \cref{app:kretschmann}). By the transformation to the coordinates chosen for \cref{eq:result-withoutz-R}, this singularity is now found at $\rma = R$ which is a finite horizon in the description of the spacetime manifold. Furthermore, we also find a coordinate singularity at $\rma = 0$, i.e. inside the horizon. A sketch of the singularities is given in \cref{fig:singularities}, giving rise to the name \textit{cylinder frame} for the line element in \cref{eq:result-withoutz-R}.

In addition to the different description of the singularities, the coefficients in the line element indicate different symmetries for an observer in the two coordinate frames. As noted in \cref{sec:cyl-vs-lp}, in \cref{eq:cyl-lp}, the coefficients of the radial ($\diff \rlp$) and axial ($\diff \zlp$) coordinates agree, whereas in \cref{eq:result-withoutz-R} this is the case for the coefficients of the radial ($\diff \rma$) and angular ($\rma \diff \varphi$) coordinates instead. 
The undistorted coordinate frame corresponds to a conformally flat spatial metric as it is the case for \cref{eq:schd-isotr-metric}, which can be expressed in cylindrical coordinates as given in \cref{eq:schd-isocyl-metric}. If the coefficients of the radial and axial coordinates differ, as found in \cref{eq:result-withoutz-R}, the surfaces of constant axial ($\zma$) coordinate are deformed but the frame remains axially symmetric. 
In contrast, if the coefficients of the radial and axial coordinates are chosen to agree, \cref{eq:cyl-lp}, the difference in the coefficients of $\diff \rlp$ and $\rlp\diff\varphi$ results in an angular distorsion of the coordinate frame. 
This frame is thus not isotropic. As a Newtonian observer is considering an isotropic coordinate frame, the cylinder frame, \cref{eq:result-withoutz-R}, is preferred in the Newtonian limit, as long as only small distances in the axial direction are considered. 

The cylinder frame has one major disadvantage being its singularity at a finite horizon $R$. This issue can be lifted by choosing coordinates where this singularity is found at $\rlp = 0$, such as the coordinate frame from \cref{eq:cyl-lp}. We thus find that at $\rma \sim R$, the coordinate frame of \cref{eq:cyl-lp} is preferred, whereas a Newtonian observer at small distances in the axial coordinate would consider the cylinder frame, \cref{eq:result-withoutz-R}.



\subsection{Approximate Solutions}\label{sec:approx}

Note that the line elements in \cref{eq:result-withoutz-R,eq:cyl-lp} are not asymptotically Minkowskian and thus do not fulfill the boundary conditions stated by \citet{quevedo1990ForPh..38..733Q}. In order to construct a metric that meets these boundary conditions and can be applied in the weak-field limit, we start from the line element \cref{eq:result-withoutz-R}. It is possible to construct approximate solutions from other line elements such as \cref{eq:cyl-lp}. However, keeping in mind that in the weak-field limit an observer would expect the metric to be undistorted in the angular direction we start from \cref{eq:result-withoutz-R}. As this solution has differing coefficients in axial and radial direction, one must be careful when considering large axial distances. Nevertheless, it is a viable approximation for a weak-field observer close to the $z = 0$ plane.
From these considerations, we find
\begin{align}
    \diff s^2 &= - \left(1 + Ce^{-\lambda \rma} \ln \frac{\rma}{R}\right) c^2 \diff t^2 + \left(1 + \frac{B e^{-\nu\rma}}{\rma^2\sqrt{C\ln \frac{\rma}{R}}}\right) \left(\diff \rma^2 + \rma^2 \diff \varphi^2\right) \breakline + \left(1 + ECe^{-\lambda\rma}\ln \frac{\rma}{R}\right) \diff \zma^2\label{eq:result-approx}.
\end{align}
This metric is not an exact solution to the \efe, but solves them approximately, as shown in \cref{app:approx}. 
By construction, this line element approaches the Minkowskian metric $\eta_{\mu\nu} = \mathrm{diag}(-1, 1, 1, 1)$ for $\rma \to \infty$, which is the weak-field limit. In the opposite case, $Ce^{-\lambda \rma} \ln \frac{\rma}{R} \gg 1$ and $\frac{B e^{-\nu\rma}}{\rma^2\sqrt{C\ln \frac{\rma}{R}}} \gg 1$,  \cref{eq:constraint-nu,eq:constraint-lambda} imply that the derivatives of $e^{-\lambda \rma}, e^{-\nu \rma}$ are relatively small in comparison and thus the line element in \cref{eq:result-approx} approaches the solution described by the metrics in \cref{eq:result-withoutz-R,eq:cyl-lp}.

Another approximate metric worth considering is 
\begin{align}
    \diff s^2 &= -\left(1 - 2 \Phi\right) c^2 \diff t^2 + \left(1 + 2 \Phi\right) \left(\diff \ran^2 + \ran^2 \diff \varphi^2 + \diff \zan^2\right)\label{eq:newt-isocyl-metric},
\end{align}
which is the Newtonian metric from \cref{eq:newt-isotr-metr} in cylindrical coordinates.
This metric constitutes the Newtonian approximation to the Schwarzschild metric, \cref{eq:schd-isotr-metric}, in cylindrical coordinates,
\begin{align}
    \diff s^2 &= -\frac{\left(1 - \Phi\right)^2}{\left(1 + \Phi\right)^2} c^2 \diff t^2 + \left(1 + \Phi\right)^4 \left(\diff \ran^2 + \ran^2 \diff \varphi^2 + \diff \zan^2\right)\label{eq:schd-isocyl-metric},
\end{align}
thus the line element in \cref{eq:newt-isocyl-metric} solves the \efe\ approximately.

The two solutions from \cref{eq:result-approx,eq:newt-isocyl-metric} differ in the choice of symmetries: In the Newtonian case, the spatial part of the metric is isotropic in all three spatial directions, while the approximate cylinder solution is constructed such that planes of constant $z$-coordinate appear isotropic (in two dimensions). This is by construction based on the line element from \ref{eq:result-withoutz-R}. It is possible to construct an approximate metric from the line element in \cref{eq:cyl-lp} or any other line element in the Lewis-Papapetrou form, if one prefers to have the coefficients of $\diff \ran$ and $\diff \zan$ in the ansatz in \cref{eq:2ansatz} to agree. However, it can be shown (see \cref{app:unique}) that the cylinder solution given by \cref{eq:result-withoutz-R} with $E = 1$ is the unique solution for which $a = h$ and $b = f$ in the ansatz in \cref{eq:2ansatz}.

Both \cref{eq:result-approx,eq:newt-isocyl-metric} give line elements for different choices of symmetries that solve the \efe\ for vanishing energy-momentum tensor. Although they are not exact solutions, they remain viable to consider as approximate metrics.

\subsection{Low-Velocity Limit}\label{sec:particles}

The Newtonian approximation, \cref{eq:newt-isocyl-metric}, is generally considered in the low-velocity limit. 
In this limit, $\diff x^i \ll c \diff t = \diff x^0$, i.e. 
\begin{align}
    -c^2\diff \tau^2 = \diff s^2 &\approx -e^{2a} c^2\diff t^2,\\
    \frac{\diff x^i}{c \diff \tau} \ll \frac{\diff x^0}{c\diff \tau} &\approx e^{-a}
\end{align}
for $\tau$ being the proper time. Hence, the geodesic equation
\begin{align}
    \frac{\diff^2 x^k}{c^2 \diff \tau^2} = -{\Gamma^k}_{\mu \nu} \frac{\diff x^\mu}{c \diff \tau} \frac{\diff x^\nu}{c \diff \tau}
\end{align}
reduces to
\begin{align}
    \frac{\diff^2 x^k}{\diff t^2} \approx e^{2a}\frac{\diff^2 x^k}{\diff \tau^2} &\approx - c^2 e^{2a}{\Gamma^k}_{00} (e^{-a})^2 = \frac{c^2}{2}\partial^k g_{00} = -\partial^k \phi\label{eq:eff-accel}
\end{align}
for an effective potential $\phi = -\frac{c^2}{2} g_{00} + \mathrm{const.}$ and can thus be compared to the classical equation of motion. However, it must be noted that the classical gradient corresponds to $\partial_k$ instead of $\partial^k$. This gives rise to an additional factor of $g^{kl}$ in \cref{eq:eff-accel}, although this change is negligible in the weak-field limit. For example, the effective potential obtained from the line element in \cref{eq:newt-isocyl-metric} is
\begin{align}
    \phi = - c^2 \Phi = -\frac{GM}{\sqrt{\ran^2 + \zan^2}} = -\frac{GM}{\risoschd}
\end{align}
which resembles the Newtonian potential.
With $\risoschd = \sqrt{\ran^2 + \zan^2}$, cf. \cref{eq:newt-isotr-metr}, The equation of motion reads
\begin{align}
    \Ddot{\risoschd} \approx -g^{\risoschd\risoschd} \partial_\risoschd \left(-\frac{GM}{\risoschd}\right) \approx -\frac{GM}{\risoschd^2}.
\end{align}
which can also be found by performing the low-velocity limit of the metric in \cref{eq:schd-isotr-metric} first and then taking the weak-field limit:
\begin{align}
    \partial_\risoschd g_{00} &= 2 \frac{1 - \frac{\Phi}{2}}{\left(1 + \frac{\Phi}{2}\right)^3} \partial_\risoschd \Phi \approx 2 \partial_\risoschd \Phi,\\
    \Ddot{\risoschd} \approx \frac{c^2}{2} g^{\risoschd\risoschd} \partial_\risoschd g_{00} &= \frac{c^2}{2} \frac{1}{\left(1 + \frac{\Phi}{2}\right)^4} \partial_\risoschd \frac{-\left(1 - \frac{\Phi}{2}\right)^2}{\left(1 + \frac{\Phi}{2}\right)^2} = \frac{c^2\partial_\risoschd \Phi}{\left(1 + \frac{\Phi}{2}\right)^4} \frac{1 - \frac{\Phi}{2}}{\left(1 + \frac{\Phi}{2}\right)^3} \approx c^2 \partial_\risoschd \Phi = -\frac{GM}{\risoschd^2}.
\end{align}
Notably, the equation of motion obtained in the low-velocity and weak-field limit from \cref{eq:schd-standard} is of the same form,
\begin{align}
    \partial_\rschd g_{00} &= 2 \partial_\rschd \Psi,\\
    \Ddot{\rschd} \approx \frac{c^2}{2} g^{\rschd\rschd} \partial_\rschd g_{00} &= \frac{c^2}{2} \left(1 - 2 \Psi\right) \partial_\rschd \left(-1 + 2 \Psi\right) = c^2 \left(1 - 2 \Psi\right) \partial_\rschd \Psi \approx c^2 \partial_\rschd \Phi = -\frac{GM}{\rschd^2}.
\end{align}
This is due to the fact that the radial coordinates $\rschd = \risoschd \left(1 + \frac{\Phi}{2}\right)^2 \to \risoschd$ agree in this case.

We can now turn to the axisymmetric metrics presented in \cref{sec:cyl-vs-lp}. For these metrics, it is \textit{a priori} not clear how the weak-field limit would have to be defined, thus we consider the low-velocity limit of the full line element. In particular, we find for the metric from \cref{eq:cyl-lp}
\begin{align}
    \partial_\rlp g_{00} &= - \frac{4 \rho^3 }{B^2},\\
    \Ddot{\rlp} \approx \frac{c^2}{2} g^{\rlp\rlp} \partial_\rlp g_{00} &= \frac{c^2}{2} \frac{B^2 C^2}{16 \rlp^4} \partial_\rlp \frac{-\rlp^4}{B^2} = -\frac{c^2C^2}{8 \rlp}.
\end{align}
In contrast to this, we have
\begin{align}
    \partial_\rma g_{00} &= - \frac{C}{\rma},\\
    \Ddot{\rma} \approx \frac{c^2}{2} g^{\rma\rma} \partial_{\rma} g_{00} &= \frac{c^2}{2} \frac{\rma^2 \sqrt{C \ln \frac{\rma}{R}}}{B} \partial_\rma \left(-C \ln \frac{\rma}{R}\right) = -\frac{c^2C\rma}{2B}  \sqrt{C \ln \frac{\rma}{R}}
\end{align}
in the coordinate frame from \cref{eq:result-withoutz-R}. In contrast to the Schwarzschild solution, we find that the equation of motion differs significantly in the two coordinate frames, as the coordinates do not agree in this case but are related by $\rlp^4 = B^2 C \ln\frac{\rma}{R}$. In both coordinate frames, the factor of $g^{kl}$ is important for the equation of motion, as we are not considering a weak-field case here.

Furthermore, consider the low-velocity limit for the approximate cylindrical solution given by \cref{eq:result-approx}. 
This line element has a well-defined weak-field limit, in which the effective potential is given by 
\begin{align}
    \phi &= \frac{Cc^2}{2}e^{-\lambda \rma} \ln \frac{\rma}{R}\label{eq:eff-pot-log}
\end{align}
and thus the equation of motion reads
\begin{align}
    \Ddot{\rma} &= - \nabla \phi = -\frac{Cc^2}{2} e^{-\lambda\rma}\frac{1 - \lambda\rma \ln\frac{\rma}{R}}{\rma}.\label{eq:eff-cutoff-accel}
\end{align}
Close to the centre, i.e. in the limit $\rma \ll \frac{1}{\lambda}$, this gives 
\begin{align}
    \Ddot{\rma} \approx \frac{Cc^2}{2 \rma} + \mathcal{O}\left(\frac{1}{\lambda}\right).\label{eq:eff-approx-accel}
\end{align}

As we are considering axisymmetric metrics here, one interesting application is to investigate rotational motion in the $z = 0$ plane. In the low-velocity limit, this can be described by 
\begin{align}
    F_{\rm central} = \frac{m v^2}{\ran} &\overset{!}{=} m |\Ddot{\ran}|, \\
    v &= \sqrt{|\Ddot{\ran}| \ran} \label{eq:flat},
\end{align}
with the rotational velocity $v$. From the derivations above, we find
\begin{align}
    v_\risoschd &\approx \sqrt{\frac{GM}{\risoschd}}, & v_\rschd &\approx \sqrt{\frac{GM}{\rschd}},
\end{align}
for the Schwarzschild solution, cf. \cref{eq:schd-standard,eq:schd-isotr-metric}, 
\begin{align}
    v_\rlp &\approx \sqrt{\frac{c^2 C^2}{8}}, & v_\rma &\approx \sqrt{\frac{c^2 C}{2B} \sqrt{C \ln \frac{\rma}{R}}},
\end{align}
for the exact axisymmetric solutions from \cref{eq:cyl-lp,eq:result-withoutz-R} and 
\begin{align}
    v_{\rm approx} &\approx \sqrt{\frac{c^2 C}{2}},
\end{align}
for the approximate metric from \cref{eq:result-approx}. Notably, the rotational velocities $v_\rlp$, corresponding to the axisymmetric solution close to the horizon, and $v_{\rm approx}$, which is applicable for a Newtonian observer, are constant with respect to the respective radial coordinate, i.e. they correspond to flat rotation curves.

\section{Discussion}\label{sec:discussion}

In this work, we present axisymmetric metrics in different coordinate frames and discuss the applicability of these frames. The solution in \cref{eq:cyl-lp} is of the Lewis-Papapetrou form and is therefore useful when comparing to other metrics. As a second choice of coordinates, we consider the cylinder frame, where the coefficients of $\diff \rma$ and $\rma \diff \varphi$ agree. 

These solutions are not suitable to describe the full spacetime of any physical object due to their limiting behaviour. Thus, to describe physical objects, we also consider the approximate solution given in \cref{eq:result-approx} derived from the cylinder solution. Notably, considering this solution in the Newtonian limit, we find the functional form of the coefficient of $c \diff t$ as a function of the radial coordinate to be crucial for the motion of a test particle. This emphasizes the importance of the choice of symmetries and the coordinate frame when comparing general relativistic solutions to observations.

\subsection{Cylinder Solution in Comparison to the Schwarzschild Metric}


Apart from considering the cylinder line element, \cref{eq:result-withoutz-R}, as the solution in \cref{eq:cyl-lp} expressed in a different coordinate frame, we can also find it from the vacuum \efe\ directly. This is presented in \cref{app:efe-cyl}. It can be derived from the same assumptions  as the Schwarzschild solution \citep{Schwarzschild1916SPAW.......189S} except for the symmetry conditions (both are static vacuum solutions). In particular, the Schwarzschild solution is spherically symmetric, while the cylinder solution possesses only axial symmetry. However, spherical symmetry also implies axial symmetry, thus the Schwarzschild solution solves the \efe\ in their form given in \cref{eq:fullrzf-g00,eq:fullrzf-g11,eq:fullrzf-g13,eq:fullrzf-g22,eq:fullrzf-g33}. This can be checked in appropriate coordinates using the metric in \cref{eq:schd-isocyl-metric}.

In this approach, the energy-momentum tensor is assumed to vanish anywhere but at the singularities, according to the vacuum \efe.
At the singularities, however, no statement can be made. This suggests for the energy-momentum tensor of the Schwarzschild solution to be of the form $\delta(\risoschd)$. Similarly, the energy-momentum tensor of the cylinder solution can be interpreted to be proportional to $\delta(\rma)$, with $\rma$ being the radial coordinate in this choice of cylindrical coordinates, but its dependence on $\zma$ is not known. 

All of the solutions considered in this work are vacuum solutions, as is the Schwarzschild solution. Although this does not agree with the physical reality, it represents an important limiting case. In situations where the mass content in the environment is negligible compared to the central mass, as it is the case for a black hole, the vacuum solution is an appropriate approximation. Similarly, the cylinder solution is a viable approximation for a cylindrical configuration where the main mass content is centered at $\rma = 0$ and the mass content in the surroundings is negligible compared to it.

\subsection{Comparison to Newtonian Gravity and Observations}

Recovering Newtonian gravity from the \efe\ by assuming vacuum, static, spherically symmetric conditions and the limits of low velocities and $\risoschd = \sqrt{\ran^2 + \zan^2} \gg R_{\mathrm{S}}$ (which corresponds to the small field limit) with the mass
assumed to be located at the singularity at $\risoschd = 0$, 
results in an effective potential $\phi = -\frac{GM}{\risoschd}$ and effective acceleration proportional to $\frac{1}{\risoschd^2}$. In contrast, applying the same conditions as in the derivation of Newtonian gravity except for the symmetry conditions, 
we find a logarithmic effective potential and accelerations proportional to $\frac{1}{\rma}$  
from the approximate cylinder line element.

A situation where rotational velocities based on Newtonian gravity, as discussed in \cref{sec:particles} are commonly considered, is the description of rotation curves in galaxies. Remarkably, the rotational velocities obtained from the Newtonian case differ from the ones observed in Nature. Newtonian gravity is not able to explain the flat rotation curves in galaxies with baryonic matter only which has been one of the reasons to introduce dark matter. Importantly, a logarithmic effective potential as obtained from the approximate cylinder solutions yields flat rotation curves when considering rotational motion in a plane of constant $z$-coordinate, without any need to introduce dark matter. 


\begin{figure}
    \centering
        \includegraphics[width=0.63\textwidth]{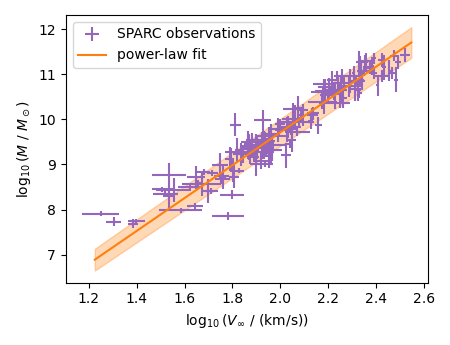}
    \caption{Fit of the velocities from the SPARC catalog \citep{SPARC:2016AJ....152..157L} to the baryonic Tully-Fisher relation, resulting in $\log_{10}\left(\frac{M}{M_\odot}\right) = (3.64 \pm 0.07) \cdot \log_{10}\left(\frac{V_\infty}{{\rm km / s}}\right) + (2.43 \pm 0.15)$. It is performed based on the mass models by \citet{SPARC:2016AJ....152..157L} and agrees within 1-2$\sigma$ with the fits in their paper. For more details see \cref{app:btfr}. 
    }
    \label{fig:btfr}
\end{figure}

In Newtonian gravity, the potential in such symmetry conditions is calculated as done in \cref{app:newt} for a line-like source. There, the source at $\ran = 0$ is assumed to be finite and of length $2a$. The cylinder solution could describe such a setting, as the energy-momentum tensor and thus the mass density is singular at $\rma = 0$ and not specified further. In the two limits discussed in \cref{app:newt}, a logarithmic potential is found for small radii and the inverse-radius law is recovered in the far-field limit. In a sense, the Newtonian treatment is interpolating between the two effective potentials. It is, however, based on the mass density distribution and only valid for a specific choice of source at $\rma = 0$. In the corresponding generally relativistic situation given by the cylinder solution, this choice of distribution is not the only valid one, as we only assume symmetry conditions and not a specific choice of mass configuration. When interpreting our solutions, we find the mass to be located at the singularities and beyond the horizon, i.e., at $\rma < R$, but no other choices are made.

People have investigated logarithmic potentials and respective accelerations to explain flat rotation curves without the help of dark matter (see \citep{Deur:2009ya,Deur2017EPJC...77..412D} for an argument based on approximately solving general relativity numerically and then enforcing a cylindrical symmetry, and \citep{deur_2020} for an analysis of the GR-SI approach involving a disk model based on a logarithmic effective potential), but here it emerges exactly and analytically from the vacuum \efe\ only based on the cylindrical symmetry. The approximate metric is only constructed to be able to describe the full spacetime of a physical object but the local metric close to the object is an exact solution.

Considering the approximate metric in the low-velocity limit, we can compare it to observations such as the SPARC catalog \citep{SPARC:2016AJ....152..157L}. This is important to fix the integration constants, such as $C$. For example, this can be done by fitting the baryonic Tully-Fisher relation (see \cref{fig:btfr,app:obs}). We find that the rotational velocity derived in \cref{sec:particles} fits the bTFR within $1\sigma$ for an acceleration scale
\begin{align}
    \Tilde{\mu} &= (1.1 \pm 0.4) \cdot 10^{-10} \frac{\mathrm{m}}{\mathrm{s}^2}\\
    C &= \sqrt{\frac{\Tilde{\mu}}{c^2}\frac{GM}{c^2}}
\end{align}
related to the integration constant.
Interestingly, this acceleration scale agrees very well with the
MOND acceleration scale \citep[see \cref{eq:MOND-scale} and][]{McGaugh2016PhRvL.117t1101M}. Intuitively, this is not surprising because both calculations are based on the SPARC data \citep{SPARC:2016AJ....152..157L}. However, the bTFR derived by MOND (\cref{app:bTFR-MOND}) differs by a factor of $4$. The large systematic error on the acceleration scale as reported by \citet{McGaugh2016PhRvL.117t1101M} allows for that. For a more thorough investigation of these factors, more data is needed. Furthermore, it is worth noticing that the results obtained from the line elements only indicate the need for a fundamental scale $\Tilde{\mu}$, the value agreeing with the MOND acceleration scale is just found by comparison to the data. In contrast to MOND, the approach taken in this work is thus not only rooted in the standard framework of GR but also more adaptive to the data.  



\subsection{Regime of Application of the Solutions}\label{sec:discuss-apply}

\begin{figure}
    \centering
    \includegraphics[width=0.6\textwidth]{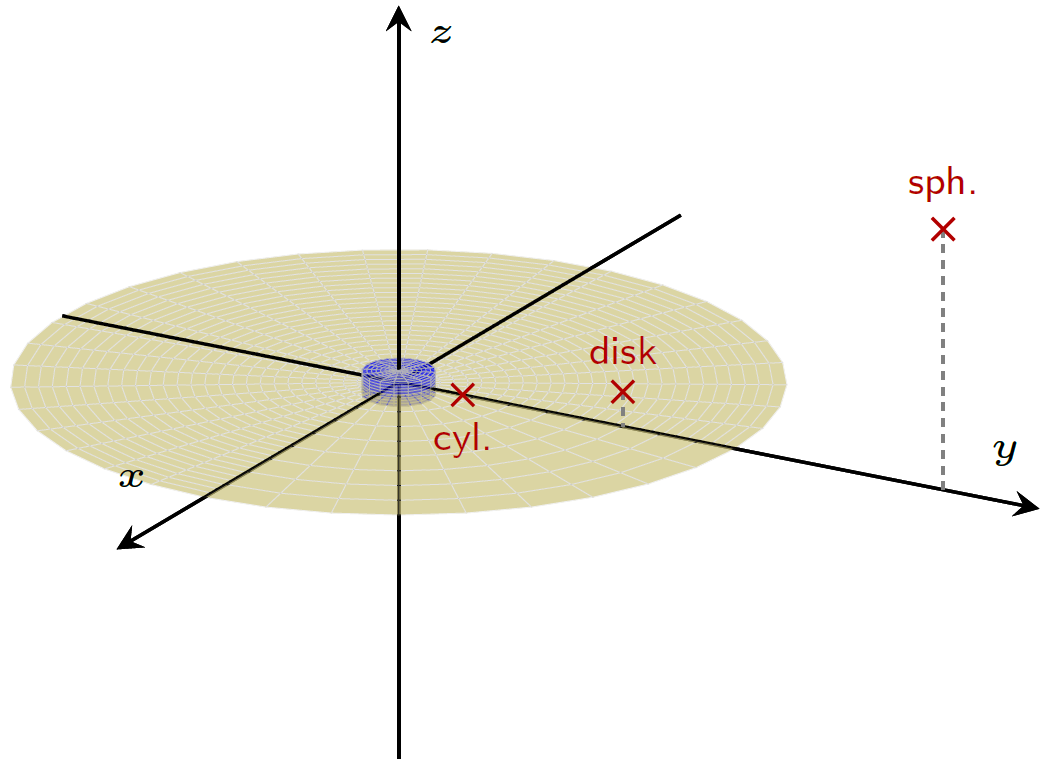}
    \caption{Sketch (not to scale) of the regimes in the environment of a galaxy where the different solutions apply (the cylindrical solution from \cref{sec:cyl-vs-lp}, the disk solution derived in \cref{app:disk} and the spherically symmetric solution from the Schwarzschild/Newtonian case).}
    \label{fig:galaxy}
\end{figure}

The difference between the line elements \cref{eq:cyl-lp,eq:result-withoutz-R} is the appearence of the local coordinate frame. In particular, the cylinder frame, \cref{eq:result-withoutz-R}, is applicable for an isotropic local observer, whereas the line element in \cref{eq:cyl-lp} is characterized by the coefficients of the radial and axial coordinates being the same. As the frame of an Newtonian observer is expected to be isotropic, the cylinder solution is preferred in this limit. 

Considering the metric in the cylinder frame as a solution of the vacuum \efe\ for the ansatz \cref{eq:2ansatz} as presented in \cref{eq:fullrzf-g00,eq:fullrzf-g11,eq:fullrzf-g13,eq:fullrzf-g22,eq:fullrzf-g33}, the cylinder solution can be viewed as a solution where the derivatives with respect to $\ran$ and $\zan$ cannot cancel each other. In particular, the cylinder solution does not involve any $\zan$-dependence at all, and a disk solution can be obtained from it by variation of constants (\cref{app:disk}). This distinguishes the cylinder solution from the Schwarzschild solution, as for the latter the terms in \cref{eq:fullrzf-g00,eq:fullrzf-g11,eq:fullrzf-g13,eq:fullrzf-g22,eq:fullrzf-g33} involving only one derivative do not vanish but it fulfills the full vacuum \efe.
Thus, the disk and cylinder line elements are not the only solutions solving the latter set of equations. However, they are the ones relevant to scenarios where the derivatives with respect to different coordinates cannot compensate each other. In terms of the potential, this means that it scales differently with different coordinates. As the approximate solution is built from the cylinder solution, this consideration applies to it as well.

The cylinder line element is however not applicable for a full description of any physical object, as it does not approach a Minkowski metric in the far-field limit. Nevertheless, it is still valid locally. In particular, as the \efe\ are differential equations and derivatives have to be taken locally, \emph{a priori} the solutions are local as well. In particular, we can imagine the cylinder solution to be valid within a disk plane at scales where no $z$-dependence is experienced and the mass content can be approximated to be located in the centre only. 
A full picture of the different regimes will involve multiple local solutions that apply in different regimes, as illustrated in \cref{fig:galaxy}. Combining these results quantitatively to describe the interpolating regimes is a non-trivial task. One approach to this is the approximate solution presented in this work (\cref{sec:approx}). 

However, for an exact form of such a model, a more involved differentiable metric or at least a different choice of coordinates would be needed. 
This solution would be an important model in the limiting case of the static vacuum spacetime, i.e. negligible and non-rotating mass distribution outside the horizon. However, lifting the assumptions of staticity to obtain a stationary metric and allowing for a non-vanishing energy-momentum tensor in the \efe\ would allow for more physical solutions \citep{Astesiano2022,Re_Galoppo:2024arXiv240303227R,Beordo:Crosta:2024MNRAS.529.4681B}.



Importantly, describing the motion of test particles with the effective potentials from \cref{sec:particles} is only possible under certain conditions. First, the physical situation has to obey the symmetry conditions of the metric, i.e., cylindrical symmetry and staticity, at least as a viable approximation. Note that stationarity is not sufficient to apply the metrics derived here. Additionally, the effective potential is derived from a vacuum metric with an energy-momentum tensor proportional to a delta distribution. We can thus only apply this effective potential to situations with negligible mass distributions outside the singularities.

Nevertheless, the line elements and their applications provide a starting point for investigating the implications of full general relativity on dark matter in galaxies. 
Notably, they do not exclude the existence of dark matter. In particular, the solutions discussed in this work still allow for cylindrically distributed dark matter. It is however impossible to directly prove any non-existence of particles. As dark matter is needed to explain further observations, it is a viable theory, if this was not the case, Occam's razor would suggest to drop this hypothesis. The results presented in this work can explain flat rotation curves in both cases.

This work shows that choosing symmetry conditions adequately is crucial when considering a gravitational system in the general relativity framework. To address the consequences of these results to their full extend, further research is needed. In particular, the regimes of applicability of these line elements and the transition between them (c.f. \cref{fig:galaxy}) have to be investigated carefully. Furthermore, the insights on the importance of symmetries should also be considered in other gravitational systems that cannot be explained with Newtonian gravity. The need for dark matter to explain observed phenomena in gravitational systems has to be re-evaluated carefully based on the individual symmetry conditions of the different systems.

\section{Conclusions}\label{sec:conclusion}


We have considered a special generally relativistic vacuum solution, the cylinder line element, in different coordinates. We find that the choice of coordinates expresses different symmetries for a local observer, thus dependent on the local situations, different coordinate frames should be chosen. In particular, a Newtonian observer should consider an isotropic coordinate frame in the low-velocity limit. 

Both the cylinder solution and the Schwarzschild solution \citep{Schwarzschild1916SPAW.......189S} describe static vacuum spacetimes but differ in the underlying symmetries. In particular, the cylinder solution depends on the cylindrical radius $\rma$ only, while the Schwarzschild solution depends on the spherical radius $\risoschd = \sqrt{\rma^2 + z^2}$ instead. 
The cylinder solution is thus applicable under the same conditions as the Schwarzschild solution and its Newtonian approximation, except for the symmetry assumptions. 
However, the asymptotic behaviour of the cylinder and Schwarzschild solutions is different, as the cylinder solution does not approach the Minkowski metric asymptotically. We thus discuss an additional approximate cylinder metric that rectifies this.

All of the solutions discussed here describe static vacuum spacetimes. They will thus not yield a full physical model for any object such as galaxies but they represent important limiting cases. In a spherically symmetric setting with a central mass being much larger than a possible additional mass configuration in the environment and for low velocities, the Schwarzschild solution reduces to the Newtonian case. This is used throughout classical mechanics very successfully and even thought to apply in the setting of galaxies. Similarly, the solutions presented here can be applied as a limiting case when studying objects with substantially high central mass, i.e. negligible mass content in the environment, and cylindrical symmetry. 
The prime example of such a situation is a disk galaxy, where the cylinder and approximate solutions are even more viable than the Schwarzschild and Newtonian line elements.

By considering the motion of particles in the low-velocity limit in the approximate line element compared to the Newtonian one, we have shown that the symmetry assumptions fundamentally change the trajectories of test particles. In particular, rotational motion in the cylindrically symmetric spacetimes yield flat rotation curves while the Schwarzschild solution reduces to Newtonian gravity in this limit. 
By fitting the results for the flat rotation curves to observations, the constants in the cylinder line element can be fixed. 
However, it must be kept in mind that the solutions cannot provide a full galaxy model as they are vacuum solutions and rely on the staticity assumption. Instead, they can only relate to the physical settings in appropriate limits and approximations. Apart from the step taken in defining the approximate line element to address the non-Minskowkian asymptotics of the exact solutions, these are further issues be solved to describe galaxies relativistically. However, the investigation of the solutions already found is an important step towards a better understanding of dark matter but cannot be seen as a full explanation of the observationally found galaxy rotation curves yet.

The fact that the symmetries considered in the observer's reference frame change the equations of motion for a particle in the respective frame provides insights into a better understanding of the influence of symmetries and coordinate frames in general relativity. 
It is crucial to investigate the regimes of applicability of these solutions to understand its implications to what is typically considered as dark matter. Additionally, the changes in the spacetime arising from the symmetries should be considered in other gravitating systems as well for a better understanding of gravity from general relativity in these situations and how it relates to dark matter. 

\section*{Acknowledgements}

I am deeply thankful to Matthias Bartelmann for his supervision and support and to Marco Galoppo for very fruitful discussions. I also thank the Astrophysics Group at Old Dominion University, namely Alexandre Deur, Bal\v sa Terzi\'c, William Clark and Emerson Rogers, for their comments and discussions on the applications and interpretation of the cylinder solution. Furthermore, I want to thank Adrian Hosak and Frederik Kortkamp for their helpful thoughts and insights over the course of developing this article.



\def\ref@jnl#1{{#1}}%
\newcommand\aj{\ref@jnl{AJ}}
\newcommand\araa{\ref@jnl{ARA\&A}}
\newcommand\apj{\ref@jnl{ApJ}}
\newcommand\apjl{\ref@jnl{ApJL}}     
\newcommand\apjs{\ref@jnl{ApJS}}
\newcommand\ao{\ref@jnl{ApOpt}}
\newcommand\apss{\ref@jnl{Ap\&SS}}
\newcommand\aap{\ref@jnl{A\&A}}
\newcommand\aapr{\ref@jnl{A\&A~Rv}}
\newcommand\aaps{\ref@jnl{A\&AS}}
\newcommand\azh{\ref@jnl{AZh}}
\newcommand\baas{\ref@jnl{BAAS}}
\newcommand\icarus{\ref@jnl{Icarus}}
\newcommand\jaavso{\ref@jnl{JAAVSO}}  
\newcommand\jrasc{\ref@jnl{JRASC}}
\newcommand\memras{\ref@jnl{MmRAS}}
\newcommand\mnras{\ref@jnl{MNRAS}}
\newcommand\pra{\ref@jnl{Phys.~Rev.~A}}
\newcommand\prb{\ref@jnl{Phys.~Rev.~B}}
\newcommand\prc{\ref@jnl{Phys.~Rev.~C}}
\newcommand\prd{\ref@jnl{Phys.~Rev.~D}}
\newcommand\pre{\ref@jnl{Phys.~Rev.~E}}
\newcommand\prl{\ref@jnl{Phys.~Rev.~Lett.}}
\newcommand\pasp{\ref@jnl{PASP}}
\newcommand\pasj{\ref@jnl{PASJ}}
\newcommand\qjras{\ref@jnl{QJRAS}}
\newcommand\skytel{\ref@jnl{S\&T}}
\newcommand\solphys{\ref@jnl{SoPh}}
\newcommand\sovast{\ref@jnl{Soviet~Ast.}}
\newcommand\ssr{\ref@jnl{SSRv}}
\newcommand\zap{\ref@jnl{ZA}}
\newcommand\nat{\ref@jnl{Nature}}
\newcommand\iaucirc{\ref@jnl{IAUC}}
\newcommand\aplett{\ref@jnl{Astrophys.~Lett.}}
\newcommand\apspr{\ref@jnl{Astrophys.~Space~Phys.~Res.}}
\newcommand\bain{\ref@jnl{BAN}}
\newcommand\fcp{\ref@jnl{FCPh}}
\newcommand\gca{\ref@jnl{GeoCoA}}
\newcommand\grl{\ref@jnl{Geophys.~Res.~Lett.}}
\newcommand\jcp{\ref@jnl{JChPh}}
\newcommand\jgr{\ref@jnl{J.~Geophys.~Res.}}
\newcommand\jqsrt{\ref@jnl{JQSRT}}
\newcommand\memsai{\ref@jnl{MmSAI}}
\newcommand\nphysa{\ref@jnl{NuPhA}}
\newcommand\physrep{\ref@jnl{PhR}}
\newcommand\physscr{\ref@jnl{PhyS}}
\newcommand\planss{\ref@jnl{Planet.~Space~Sci.}}
\newcommand\procspie{\ref@jnl{Proc.~SPIE}}

\newcommand\actaa{\ref@jnl{AcA}}
\newcommand\caa{\ref@jnl{ChA\&A}}
\newcommand\cjaa{\ref@jnl{ChJA\&A}}
\newcommand\jcap{\ref@jnl{JCAP}}
\newcommand\na{\ref@jnl{NewA}}
\newcommand\nar{\ref@jnl{NewAR}}
\newcommand\pasa{\ref@jnl{PASA}}
\newcommand\rmxaa{\ref@jnl{RMxAA}}

\newcommand\maps{\ref@jnl{M\&PS}}
\newcommand\aas{\ref@jnl{AAS Meeting Abstracts}}
\newcommand\dps{\ref@jnl{AAS/DPS Meeting Abstracts}}

\let\astap=\aap 
\let\apjlett=\apjl 
\let\apjsupp=\apjs 
\let\applopt=\ao 

\bibliography{refs_update}







\appendix
\counterwithin{equation}{section}
\counterwithin{figure}{section}

\section{Einstein Field Equations for the Axisymmetric Ansatz}\label{app:efe-cyl}
\newcommand{\neweqlineindent}{\nonumber\\&\ \ \ \ \ \ \ }

\subsection{Vacuum Einstein Field Equations}\label{app:efe}

Given the ansatz from \cref{eq:2ansatz}, we can choose the dual Cartan tetrad
\begin{align}
    \theta^0 &= e^a \diff t,
    &\theta^1 &= e^b \diff \ran,
    &\theta^2 &= e^f \ran \diff \varphi,
    &\theta^3 &= e^h \diff \zan\label{eq:tetrad},
\end{align}
such that $\diff s^2 = \eta_{\mu\nu} \theta^\mu \theta^\nu$ for $\eta_{\mu\nu} = \mathrm{diag}(-1, 1, 1, 1)$ the Minkowski metric. Following the Cartan formalism, we can derive the connection 1-forms, ${\omega^\mu}_\nu = {\Gamma^\mu}_{\alpha \nu} \diff x^\alpha$, that relate to the Christoffel symbols, and the curvature 2-forms ${\Omega^\mu}_\nu$ using Cartan's structure equations \citep{Cartan1923} for the torsion-free Levi-Civita connection (i.e. torsion 2-forms $\Theta^\mu = 0$),
\begin{align}
    0 = \Theta^\mu &= \diff \theta^\mu + {\omega^\mu}_\nu \wedge \theta^\nu\label{eq:cartan-torsion},\\
    {\Omega^\mu}_\nu &= \diff {\omega^\mu}_\nu + {\omega^\mu}_\lambda \wedge {\omega^\lambda}_\nu\label{eq:cartan-curv}.
\end{align}
We then find
\begin{align}
    {\omega^0}_1 &= a_{,\ran} e^{-b} \theta^0 &&= {\omega^1}_0,\\
    {\omega^0}_3 &= a_{,\zan} e^{-h} \theta^0 &&= {\omega^3}_0 ,\\
    {\omega^1}_3 &= b_{,\zan} e^{-h} \theta^1 - h_{,\ran} e^{-b} \theta^3 &&= - {\omega^3}_1 ,\\
    {\omega^2}_1 &= \left(f_{,\ran} + \frac{1}{\ran}\right) e^{-b} \theta^2 &&= -{\omega^1}_2,\\
    {\omega^2}_3 &= f_{,\zan} e^{-h} \theta^2 &&= -{\omega^3}_2,
\end{align}
and 
\begin{align}
    {\Omega^0}_1(\mathfrak{e}_1, \mathfrak{e}_0) &= e^{-2b}(a_{,\ran\ran} - a_{,\ran}b_{,\ran} + {a_{,\ran}}^2) + a_{,\zan}b_{,\zan}e^{-2h},\label{eq:om01}\\
    {\Omega^0}_1(\mathfrak{e}_3, \mathfrak{e}_0) &= e^{-h-b}(a_{,\ran \zan} - a_{,\ran}b_{,\zan} + a_{,\ran}a_{,\zan} - a_{,\zan} h_{,\ran}), \label{eq:om013}\\
    {\Omega^0}_3(\mathfrak{e}_3, \mathfrak{e}_0) &= e^{-2h}(a_{,\zan\zan} - a_{,\zan}h_{,\zan} + {a_{,\zan}}^2) + a_{,\ran} h_{,\ran} e^{-2b},\label{eq:om03}\\
    {\Omega^0}_3(\mathfrak{e}_1, \mathfrak{e}_0) &= e^{-h-b}(a_{,\zan\ran} - a_{,\zan}h_{,\ran} + a_{,\zan}a_{,\ran} - a_{,\ran}b_{,\zan}),\label{eq:om031}\\
    {\Omega^2}_1(\mathfrak{e}_1, \mathfrak{e}_2) &= e^{-2b}\left(f_{,\ran\ran} - \frac{\ran f_{,\ran} + 1}{\ran}b_{,\ran} + {f_{,\ran}}^2 + 2\frac{f_{,\ran}}{\ran}\right) + b_{,\zan}f_{,\zan}e^{-2h}\label{eq:om21}\\
    {\Omega^2}_1(\mathfrak{e}_3, \mathfrak{e}_2) &= e^{-h-b}\left(f_{,\ran \zan} - \frac{\ran f_{,\ran} + 1}{\ran}b_{,\zan} + \frac{\ran f_{,\ran} + 1}{\ran}f_{,\zan} - f_{,\zan} h_{,\ran}\right),\label{eq:om213}\\
    {\Omega^1}_3(\mathfrak{e}_3, \mathfrak{e}_1) &= e^{-2h}(b_{,\zan\zan} - b_{,\zan}h_{,\zan} + {b_{,\zan}}^2) + e^{-2b}(h_{,\ran\ran} - b_{,\ran}h_{,\ran} + {h_{,\ran}}^2),\label{eq:om13}\\
    {\Omega^0}_2(\mathfrak{e}_0, \mathfrak{e}_2) &= - \left(a_{,\ran}\left(f_{,\ran} + \frac{1}{\ran}\right) e^{-2b} + a_{,\zan}f_{,\zan} e^{-2h}\right),\label{eq:om02}\\
    {\Omega^2}_3(\mathfrak{e}_1, \mathfrak{e}_2) &= e^{-h-b}\left(f_{,\zan\ran} - f_{,\zan}h_{,\ran} + (f_{,\zan} - b_{,\zan})\left(f_{,\ran} + \frac{1}{\ran}\right)\right),\label{eq:om231}\\
    {\Omega^2}_3(\mathfrak{e}_3, \mathfrak{e}_2) &= e^{-2h}(f_{,\zan\zan} - f_{,\zan}h_{,\zan} + {f_{,\zan}}^2) + h_{,\ran}\left(f_{,\ran} + \frac{1}{\ran}\right) e^{-2b}.\label{eq:om23}
\end{align}
where $\mathfrak{e}_\mu$ are the covariant dual vectors to the contravariant tetrad $\theta^\mu$, i.e. $\theta^\mu(\mathfrak{e}_\nu) = \delta^\mu_\nu$.
As the curvature two-forms correspond to the Riemann tensor by ${\Omega^\mu}_\nu = {R^\mu}_{\nu\alpha\beta} \theta^\alpha \wedge \theta^\beta$, we can use this to derive the Einstein tensor $G_\mu\nu$ and thus the \efe:
\begin{align}
    0 = G_{00} &= -e^{-2b}\left(f_{,\ran\ran} + \frac{f_{,\ran}}{\ran} + (f_{,\ran} - b_{,\ran})\left(f_{\ran}+ \frac{1}{\ran}\right) + \frac{h_{,\ran}}{\ran} + h_{,\ran}(f_{,\ran} - b_{,\ran}) + h_{,\ran\ran} + {h_{,\ran}}^2\right)\neweqlineindent- e^{-2h} \left({b_{,\zan}}^2 + {f_{,\zan}}^2 + {b_{,\zan}}f_{,\zan} + b_{,\zan\zan} + f_{,\zan\zan} - b_{,\zan}h_{,\zan} -f_{,\zan}h_{,\zan}\right)\label{eq:fullrzf-g00},\\
    0 = G_{11} &= e^{-2b}\left(a_{,\ran}f_{,\ran} + \frac{a_{,\ran}}{\ran} + a_{,\ran} h_{,\ran} + h_{,\ran}f_{,\ran} + \frac{h_{,\ran}}{\ran}\right)\neweqlineindent
    + e^{-2h} \left(a_{,\zan}f_{,\zan} + a_{,\zan\zan} - a_{,\zan}h_{,\zan} + {a_{,\zan}}^2 + f_{,\zan\zan} - f_{,\zan}h_{,\zan} + {f_{,\zan}}^2\right)\label{eq:fullrzf-g11},\\
    0 = G_{22} &= e^{-2b}\left(a_{,\ran\ran} - a_{,\ran}b_{,\ran} + {a_{,\ran}}^2 + a_{,\ran} h_{,\ran} + h_{,\ran\ran} - b_{,\ran}h_{,\ran} + {h_{,\ran}}^2\right)\neweqlineindent  
    + e^{-2h} \left(a_{,\zan}b_{,\zan} + a_{,\zan\zan} - a_{,\zan}h_{,\zan} + {a_{,\zan}}^2 + b_{,\zan\zan} - b_{,\zan}h_{,\zan} + {b_{,\zan}}^2\right)\label{eq:fullrzf-g22},\\
    0 = G_{33} &= e^{-2b}\left(a_{,\ran\ran} + {a_{,\ran}}^2 + a_{,\ran}(f_{,\ran} - b_{,\ran}) + \frac{a_{,\ran}}{\ran} + f_{,\ran\ran} + \frac{f_{,\ran}}{\ran} + (f_{,\ran} - b_{,\ran})\left(f_{\ran}+ \frac{1}{\ran}\right)\right)\neweqlineindent  + e^{-2h}\left(a_{,\zan}(b_{,\zan} + f_{,\zan}) + {b_{,\zan}}f_{,\zan}\right)\label{eq:fullrzf-g33},\\
    0 = G_{13} &= -e^{-b-h} \left(a_{,\ran \zan} - a_{,\ran}b_{,\zan} + a_{,\ran}a_{,\zan} - a_{,\zan} h_{,\ran} + b_{,\ran \zan} - b_{,\zan} h_{,\ran} + (f_{,\zan} - b_{,\zan})\left(f_{\ran}+ \frac{1}{\ran}\right)\right)\label{eq:fullrzf-g13}.
\end{align}

These equations 
are solved by various metrics such as the Schwarzschild solution in cylindrical coordinates, \cref{eq:schd-isocyl-metric}, for $b = h = f$, the line element in \cref{eq:cyl-lp} for $b = h \neq f$ in the coordinates $(\ran, \zan) = (\rlp, \zlp)$ and the cylinder line element, \cref{eq:result-withoutz-R}, for $b = f \neq h$ in the coordinates $(\ran, \zan) = (\rma, \zma)$. In the latter case we also obtain $a_{,\rma} = h_{,\rma}$, thus the functions $a$ and $h$ only differ by a constant, which can be absorbed into $\diff \zma$ by rescaling. 

\subsection{Kretschmann Scalar}\label{app:kretschmann}

We can furthermore derive the Kretschmann scalar 
\begin{align}
    \mathcal{K} = \Bar{R}_{abcd} \Bar{R}^{abcd}\label{eq:def-kretschmann}
\end{align}
In the case of $b = f$ in the ansatz \cref{eq:2ansatz}, as it is the case for the line element \cref{eq:result-withoutz-R}, we find 
\begin{align}
    -2 {{R}^0}_{110} = -2 {{R}^1}_{331} = 2 {{R}^0}_{202} = -2 {{R}^2}_{332} = {{R}^0}_{330} = {{R}^2}_{112} &= \left(\partial_\rma a\right)^2
\end{align}
and thus
\begin{align}
    \mathcal{K} 
    &= 12 ({g^{\rma\rma}})^2(\partial_\rma a)^4
    = \frac{3 C}{4 B^2 \ln^3\frac{\rma}{R}}.
\end{align}
In the coordinates of the line element \cref{eq:cyl-lp}, i.e. using the coordinate transformation given by \cref{eq:coordchange}, we find
\begin{align}
    \mathcal{K} 
    &= \frac{3 C^4 B^4}{4 \rlp^{12}}.
\end{align}
This implies that the metric has a curvature singularity at $\rma = R$ which corresponds to $\rlp = 0$. For comparison, we note that the Kretschmann scalar for the Schwarzschild solution, \cref{eq:schd-standard}, is \citep[e.g.,][]{henry_kretschmann2000ApJ...535..350H}
\begin{align}
    \mathcal{K} = \frac{48 G^2 M^2}{c^4\rschd^6}.
\end{align}


\subsection{Uniqueness of the Cylinder Solution}\label{app:unique}

In the case $a = h$ and $b = f$ in the ansatz in \cref{eq:2ansatz}, the \efe\ in \cref{eq:fullrzf-g00,eq:fullrzf-g11,eq:fullrzf-g13,eq:fullrzf-g22,eq:fullrzf-g33} reduce to 
\begin{align}
    0 &= b_{,\rma\rma} + \frac{b_{,\rma}}{\rma} + \frac{a_{,\rma}}{\rma} + a_{,\rma\rma} + {a_{,\rma}}^2,\label{eq:ah-derivb}\\
    0 &= 2a_{,\rma}b_{,\rma} + 2\frac{a_{,\rma}}{\rma} + {a_{,\rma}}^2,\label{eq:ah-prodb}\\
    0 &= 2a_{,\rma\rma} - 2a_{,\rma}b_{,\rma} + 3{a_{,\rma}}^2.\label{eq:ah-cancelb}
\end{align}

For $a_{,\rma} = 0$, \cref{eq:ah-prodb,eq:ah-cancelb} are trivially fulfilled and the set of equations has a unique solution $(a, b) = (\mathrm{const.}, b)$ with $b$ the solution to $b_{,\rma\rma} = -\frac{b_{,\rma}}{\rma}$ (this solution is unique according to the Picard-Lindelöf theorem as the right-hand side is continuous in $\rma$ and Lipschitz continuous in $b_{,\rma}$).

Considering the case $a_{,\rma} \neq 0$, \cref{eq:ah-derivb,eq:ah-prodb,eq:ah-cancelb} are equivalent to
\begin{align}
    0 &= 2b_{,\rma\rma} + 2\frac{b_{,\rma}}{\rma} + 2\frac{a_{,\rma}}{\rma} + 2a_{,\rma\rma} + 2{a_{,\rma}}^2,\label{eq:ah-stillb}\\
    0 &= 2b_{,\rma} + 2\frac{1}{\rma} + {a_{,\rma}},\label{eq:ah-defb}\\
    0 &= 2a_{,\rma\rma} + 2\frac{a_{,\rma}}{\rma} + 4{a_{,\rma}}^2.\label{eq:ah-nob}
\end{align}
which, by taking the derivative of \cref{eq:ah-defb} reduces to
\begin{align}
    0 &= 2b_{,\rma} + \frac{2}{\rma} + {a_{,\rma}},\\
    0 &= \frac{a_{,\rma}}{\rma} + a_{,\rma\rma} + 2{a_{,\rma}}^2.
\end{align}
This can be reformulated as first order differential equation for the vector $y = (b, a, k)^T$:
\begin{align}
    \frac{\partial}{\partial \rma} y =\frac{\partial}{\partial \rma}\left(\begin{array}{c}
        b \\ a \\
        k
    \end{array}\right) = \left(\begin{array}{c}
        -\frac{1}{\rma} - \frac{k}{2} \\ k \\
        -\frac{k}{\rma} - 2{k}^2\end{array}\right) = f(\rma, y).
\end{align}
The components of $f$ are continuous in $\rma$ and as they are polynomials in $k$, they are Lipschitz continuous in $k$ on an interval $[k_1, k_2]$. On this interval, we can thus apply the Picard-Lindelöf theorem implying that the solution to this differential equation is unique. Solving the equation yields
\begin{align}
    y = \left(\begin{array}{c}
        b \\ a \\
        k
    \end{array}\right) = \left(\begin{array}{c}
        -\frac{1}{4}\ln \left(C\ln \frac{\rma}{R}\right) - \ln \rma + \frac{1}{2} \ln B \\
        \frac{1}{2}\ln (C \ln \frac{\rma}{R}) \\
        \frac{1}{2 \rma \ln \frac{\rma}{R}}
    \end{array}\right)
\end{align}
in this interval $[k_1, k_2]$ which corresponds to an interval $[\rma_2, \rma_1]$ for $0 < R < \rma_2 < \rma_1 < \infty$. The solution found in this interval can be extended up to the singularity, giving a unique solution on $(R, \infty) \supset [\rma_2, \rma_1]$.

\subsection{Approximate Solution}\label{app:approx}

To verify that the line element \cref{eq:result-approx} solves the vacuum \efe\ approximately, consider
\begin{align}
    e^{2a} &= 1 + e^{2 \Tilde{a}(\rma)} e^{-\lambda \rma},\label{eq:approx-a}\\
    e^{2b} = e^{2f} &= 1 + e^{2 \Tilde{b}(\rma)} e^{-\nu \rma},\label{eq:approx-b}\\
    e^{2h} &= 1 + e^{2 \Tilde{h}(\rma)} e^{-\sigma \rma},\label{eq:approx-c}
\end{align}
for constant scales $\lambda, \nu, \sigma$.
With these definitions, \cref{eq:result-approx} can be written in the form of \cref{eq:2ansatz} by choosing $e^{2 \Tilde{a}(\rma)}, e^{2 \Tilde{b}(\rma)} = e^{2 \Tilde{f}(\rma)}, e^{2 \Tilde{h}(\rma)}$ to be the respective coefficients of the line element in \cref{eq:result-withoutz-R}.
Note that $a, b, f, h$ have to be non-negative as the exponential functions on the right hand sides of the equations give positive values. For $\Tilde{a}, \Tilde{b}, \Tilde{h}$, this can be confirmed from \cref{eq:result-withoutz-R} for $\rma > R$. Thus, $e^{2a} > 1$, which implies $a > 0$, and the same applies to $b$ and $h$. 
Such a metric is asymptotically Minkowskian by construction but does not solve the vacuum \efe\ exactly. It can be shown that for $a_{,\rma} = h_{,\rma}$, i.e. for $\lambda = \sigma$ as $\Tilde{a}_{,\rma} = \Tilde{h}_{,\rma}$ holds for \cref{eq:result-withoutz-R}, and $a_{,\zma} = h_{,\zma} = b_{\zma} = 0$, the vacuum \efe, \cref{eq:fullrzf-g00,eq:fullrzf-g11,eq:fullrzf-g13,eq:fullrzf-g22,eq:fullrzf-g33}, reduce to
\begin{align}
    0 &= 2a_{,\rma} b_{,\rma} + 2 \frac{a_{\rma}}{\rma} + {a_{,\rma}}^2\label{eq:killz-4th-eq},\\
    0 &= 2\frac{a_{,\rma}}{\rma} + 4{a_{,\rma}}^2 + 2a_{,\rma\rma}\label{eq:killz-aa},\\
    0 &= b_{,\rma\rma} + \frac{b_{,\rma}}{\rma} - {a_{,\rma}}^2\label{eq:killz-ba-red}.
\end{align}
Inserting \cref{eq:approx-a,eq:approx-b,eq:approx-c} yields
\begin{align}
    0 &= 2\left(\Tilde{a}_{,\rma} - \lambda\right) \frac{e^{2 \Tilde{a}} e^{-\lambda\rma}}{1 + e^{2 \Tilde{a}} e^{-\lambda\rma}} \left(\Tilde{b}_{,\rma} - \nu\right) \frac{e^{2 \Tilde{b}} e^{-\nu\rma}}{1 + e^{2 \Tilde{b}} e^{-\nu\rma}} + 2 \left(\frac{\Tilde{a}_{,\rma}}{\rma} - \frac{\lambda}{\rma}\right) \frac{e^{2 \Tilde{a}} e^{-\lambda\rma}}{1 + e^{2 \Tilde{a}} e^{-\lambda\rma}} \breakline + \left({\Tilde{a}_{,\rma}}^2 - 2 \Tilde{a}_{,\rma}\lambda + \lambda^2\right) \left(\frac{e^{2 \Tilde{a}} e^{-\lambda\rma}}{1 + e^{2 \Tilde{a}} e^{-\lambda\rma}}\right)^2\label{eq:approx-efe-4th-eq},\\
    0 &= \left(\frac{\Tilde{a}_{,\rma}}{\rma} - \frac{\lambda}{\rma}\right) \frac{e^{2 \Tilde{a}} e^{-\lambda\rma}}{1 + e^{2 \Tilde{a}} e^{-\lambda\rma}} + 2\left({\Tilde{a}_{,\rma}}^2 - 2 \Tilde{a}_{,\rma}\lambda + \lambda^2\right) \left(\frac{e^{2 \Tilde{a}} e^{-\lambda\rma}}{1 + e^{2 \Tilde{a}} e^{-\lambda\rma}}\right)^2 \breakline + \Tilde{a}_{,\rma\rma} \frac{e^{2 \Tilde{a}} e^{-\lambda\rma}}{1 + e^{2 \Tilde{a}} e^{-\lambda\rma}} + \boxed{\left(\Tilde{a}_{,\rma} - \lambda\right)^2 \frac{e^{2 \Tilde{a}} e^{-\lambda\rma}}{(1 + e^{2 \Tilde{a}} e^{-\lambda\rma})^2}}\label{eq:approx-efe-aa},\\
    0 &= \Tilde{b}_{,\rma\rma}\frac{e^{2 \Tilde{b}} e^{-\nu\rma}}{1 + e^{2 \Tilde{b}} e^{-\nu\rma}} + \boxed{\left(\Tilde{b}_{,\rma} - \nu\right)^2 \frac{e^{2 \Tilde{b}} e^{-\nu\rma}}{(1 + e^{2 \Tilde{b}} e^{-\nu\rma})^2}} + \left(\frac{\Tilde{b}_{,\rma}}{\rma} - \frac{\nu}{\rma}\right) \frac{e^{2 \Tilde{b}} e^{-\nu\rma}}{1 + e^{2 \Tilde{b}} e^{-\nu\rma}} \breakline - \left({\Tilde{a}_{,\rma}}^2 - 2 \Tilde{a}_{,\rma}\lambda + \lambda^2\right) \left(\frac{e^{2 \Tilde{a}} e^{-\lambda\rma}}{1 + e^{2 \Tilde{a}} e^{-\lambda\rma}}\right)^2\label{eq:approx-efe-ba-red}.
\end{align}

In the limit $e^{2 \Tilde{a}} e^{-\lambda\rma}, e^{2 \Tilde{b}} e^{-\nu\rma} \ll 1$, i.e. the far field case where the metric becomes Minkowskian, these equations are fulfilled by construction. This is the case for any choice of $\lambda$ and $\nu$ for sufficiently high values of $\rma$. 

Conversely, when considering $e^{2 \Tilde{a}} e^{-\lambda\rma}, e^{2 \Tilde{b}} e^{-\nu\rma} \gg 1$, we also have $e^{2 \Tilde{a}}, e^{2 \Tilde{b}} \gg 1$ due to $e^{-\lambda\rma}, e^{-\nu\rma} \leq 1$. Choosing the scales $\lambda, \nu$ such that 
\begin{align}
    \lambda &\ll |\Tilde{a}_{,\rma}| = \frac{1}{2 \rma \ln \frac{\rma}{R}},\label{eq:constraint-lambda}\\
    \nu &\ll |\Tilde{b}_{,\rma}| = \frac{1}{4 \rma \ln \frac{\rma}{R}} + \rma.\label{eq:constraint-nu}
\end{align}
we find by using \cref{eq:killz-4th-eq,eq:killz-aa,eq:killz-ba-red} that the only remaining terms in \cref{eq:approx-efe-aa,eq:approx-efe-4th-eq,eq:approx-efe-ba-red} are the boxed ones. We can apply a Taylor expansion in $\lambda$ and $\nu$, respectively,
\begin{align}
    \frac{e^{2 \Tilde{a}} e^{-\lambda\rma}}{(1 + e^{2 \Tilde{a}} e^{-\lambda\rma})^2} &\approx \frac{e^{2 \Tilde{a}}}{(1 + e^{2 \Tilde{a}})^2} \ll 1 \\ 
    \frac{e^{2 \Tilde{b}} e^{-\nu\rma}}{(1+ e^{2 \Tilde{b}} e^{-\nu\rma})^2} &\approx \frac{e^{2 \Tilde{b}}}{(1 + e^{2 \Tilde{b}})^2} \ll 1 
\end{align}
and thus these terms are negligible as well.

In the intermediate regime, we have 
\begin{align}
    e^{2 \Tilde{a}} e^{-\lambda\rma} &\approx 1,\\
    e^{2 \Tilde{a}} &\approx e^{\lambda \rma},\\
    a_{,\rma} &\approx \lambda = \mathrm{const.}
\end{align}
and 
\begin{align}
    e^{2 \Tilde{b}} e^{-\nu\rma} &\approx 1,\\
    e^{2 \Tilde{b}} &\approx e^{\nu \rma},\\
    b_{,\rma} &\approx \nu = \mathrm{const.}
\end{align}
and thus \cref{eq:approx-efe-4th-eq,eq:approx-efe-ba-red,eq:approx-efe-aa} are fulfilled as well.
This proves that the line element \cref{eq:result-approx} approximately solves the \efe. It is furthermore asymptotically Minkowski by construction.

In terms of the Ernst equation \citep{Ernst1968PhRv..167.1175E,quevedo1990ForPh..38..733Q}, the well-behavedness of the solution can be verified by considering the coordinate transformation 
\begin{align}
    \rlp &= L \rma,\\
    L &= \sqrt{\left(1 + Ce^{-\lambda \rma} \ln \frac{\rma}{R}\right) \left(1 + \frac{B e^{-\nu\rma}}{\rma^2\sqrt{C\ln \frac{\rma}{R}}}\right)},\\
    \zlp &= \sqrt{\frac{1 + ECe^{-\lambda \rma} \ln \frac{\rma}{R}}{1 + \frac{B e^{-\nu\rma}}{\rma^2\sqrt{C\ln \frac{\rma}{R}}}}}\left(\rma \frac{\partial L}{\partial \rma} + L\right)\zma.
\end{align}
resulting in
\begin{align}
    \diff s^2 &= - \left(1 + Ce^{-\lambda \rma} \ln \frac{\rma}{R}\right) c^2 \diff t^2 + \frac{1}{1 + Ce^{-\lambda \rma} \ln \frac{\rma}{R}} \left(\left(\frac{\rma}{L} \frac{\partial L}{\partial \rma} + 1\right)^{-2}(\diff \rlp^2 + \diff \zlp^2) + \rlp^2 \diff \varphi^2\right).
\end{align}
This line element not only solves the Ernst \cref{eq:ernst} for the Ernst potential $E = 1 + Ce^{-\lambda \rma(\rlp)} \ln \frac{\rma(\rlp)}{R}$ approximately but also fulfills the boundary conditions \citep{quevedo1990ForPh..38..733Q} for $0 < \nu < \lambda < 2 \nu$. Note that these additional constraints on $\lambda, \nu$ do not contradict \cref{eq:constraint-lambda,eq:constraint-nu} found by demaning the line element to approximately solve the \efe.

\subsection{Disk Solution}\label{app:disk}

Starting from the cylinder line element \cref{eq:result-withoutz-R}, or equivalently 
\begin{align}
    \diff s^2 &= - \left(C\ln \frac{\rma}{R}\right) c^2 \diff t^2 + \frac{1}{\rma^2\sqrt{\Tilde{C} \ln \frac{\rma}{R}}} \left(\diff \rma^2 + \rma^2 \diff \varphi^2\right) + E\left(C\ln \frac{\rma}{R}\right) \diff \zma^2
\end{align}
for $\Tilde{C} = \frac{C}{B^2}$, we can introduce $\zma$-dependencies by varying the constants $R, C, E, \Tilde{C}$ with respect to $\zma$. For the functions $a, b, h$ in the line element (cf. \cref{eq:2ansatz})
\begin{align}
    a &= \frac{1}{2}\ln (C \ln \rho + D)\label{eq:killz-a},\\
    b &= -\frac{1}{4}\ln (C \ln \rho + D) - \ln \rho + \frac{1}{2} \ln B\label{eq:killz-b},\\
    h &= a + \frac{1}{2} \ln E\label{eq:killz-c},
\end{align}
the terms in \cref{eq:fullrzf-g00,eq:fullrzf-g11,eq:fullrzf-g22,eq:fullrzf-g33} involving derivatives with respect to $\rma$ only combine to zero both with or without $\zma$-dependence in the constants. Thus, the equations to be solved by the ($\zma$-dependent) coefficients in \cref{eq:result-withoutz-R} are
\begin{align}
    0 &= 3{b_{,\zma}}^2 + 2b_{,\zma\zma} - 2b_{,\zma}h_{,\zma},\label{eq:havez-g00}\\
    0 &= a_{,\zma}b_{,\zma} + a_{,\zma\zma} - a_{,\zma}h_{,\zma} + {a_{,\zma}}^2 + b_{,\zma\zma} - b_{,\zma}h_{,\zma} + {b_{,\zma}}^2\label{eq:havez-g11},\\
    0 &= a_{,\zma}b_{,\zma} + a_{,\zma\zma} - a_{,\zma}h_{,\zma} + {a_{,\zma}}^2 + b_{,\zma\zma} - b_{,\zma}h_{,\zma} + {b_{,\zma}}^2\label{eq:havez-g22},\\
    0 &= 2a_{,\zma}b_{,\zma} + {b_{,\zma}}^2\label{eq:havez-g33},\\
    0 &= a_{,\rma \zma} - a_{,\rma}b_{,\zma} + a_{,\rma}a_{,\zma} - a_{,\zma} h_{,\rma} + b_{,\rma \zma} - b_{,\zma} h_{,\rma}.\label{eq:havez-g13}
\end{align}
Furthermore, these coefficients also solve \cref{eq:killz-4th-eq} and thus
\begin{align}
    a_{,\rma} = -2b_{,\rma} - \frac{2}{\rma}
\end{align}
which together with $a_{,\rma} = h_{,\rma}$ reduces \cref{eq:havez-g13} to
\begin{align}
    0 &= -b_{,\rma \zma} + 4b_{,\rma}b_{,\zma} + \frac{4b_{,\zma}}{\rma}.
\end{align}
For $b$ given by \cref{eq:killz-b} with $\zma$-dependent $R$ and $\Tilde{C} = \frac{C}{B^2}$, this implies
\begin{align}
    0 &= \frac{1}{4\rma R} \frac{1}{\ln^2\frac{\rma}{R}}R_{,\zma} + \frac{4\left(\Tilde{C}_{,\zma} \ln \frac{\rma}{R} - \frac{\Tilde{C} R_{,\zma}}{R}\right)}{16\rma\Tilde{C}\ln\frac{\rma}{R}}\left(\frac{1}{\ln \frac{\rma}{R}} + 1\right) - \frac{4\left(\Tilde{C}_{,\zma} \ln \frac{\rma}{R} - \frac{\Tilde{C} R_{,\zma}}{R}\right)}{16\rma \Tilde{C} \ln\frac{\rma}{R}}\\
    &= \frac{\Tilde{C}_{,\zma}}{4 \rma \Tilde{C}} \frac{1}{\ln\frac{\rma}{R}}
\end{align}
and thus $\Tilde{C}_{,\zma} = 0$. 

Let us now assume $b_{,\zma} \neq 0$, which by $\Tilde{C}_{,\zma} = 0$ implies $R_{,\zma} \neq 0$. Then, \cref{eq:havez-g33} implies $b_{,\zma} = -2 a_{,\zma}$ and by varying the constant $E$ in the $h$ function given in \cref{eq:killz-c}, we find $h_{,\zma} = a_{,\zma} + \frac{E_{,\zma}}{2E}$. Inserting this into \cref{eq:havez-g00}, we obtain
\begin{align}
    0 &= \left(4{b_{,\zma}} + 2\frac{b_{,\zma\zma}}{b_{,\zma}}\right)b_{,\zma} - \frac{E_{,\zma}}{E} b_{,\zma},\\
    &= \left(\frac{1}{R} \frac{1}{\ln\frac{\rma}{R}} R_{,\zma} - \frac{2 R_{,\zma}}{R} + \frac{2 R_{,\zma\zma}}{R_{,\zma}} + \frac{2 R_{,\zma}}{R \ln\frac{\rma}{R}}\right)b_{,\zma} - \frac{E_{,\zma}}{E}b_{,\zma} . 
\end{align}
As both $R$ and $E$ are independent of $\rma$, this is only possible for $b_{,\zma} = 0$ which contradicts the assumption. We conclude that $b_{,\zma} = 0$ and as $\Tilde{C}_{,\zma} = 0$, this implies $R_{,\zma} = 0$. However, note that with these conclusions $E_{,\zma}$ does \textit{not} have to vanish for the above equation to be fulfilled.

As we have shown that $b_{,\zma} = 0$ and using $a_{,\rma} = h_{,\rma}$ from \cref{eq:killz-a,eq:killz-c}, the \cref{eq:havez-g00,eq:havez-g11,eq:havez-g22,eq:havez-g33,eq:havez-g13} reduce to
\begin{align}
    0 &= a_{,\zma\zma} - a_{,\zma}h_{,\zma} + {a_{,\zma}}^2\label{eq:havez-g11-red},\\
    0 &= a_{,\rma \zma}\label{eq:havez-g13-red}.
\end{align}
Thus, additional terms added to \cref{eq:killz-a} can depend on $\zma$. Such terms then correspond to factors $\epsilon(\zma)$ in the coefficient $e^{2a}$, i.e. $e^{2a} = \left.e^{2a}\right|_{\rm cyl.} \epsilon(\zma)$ where the subscript $_{\rm cyl.}$ refers to the corresponding term in the cylinder solution which is independent of $\zma$. The $\zma$-dependence in the $h$ function follows from \cref{eq:havez-g11-red} which is solved by
\begin{align}
    h &= a + \ln a_{,\zma} + \frac{1}{2}\ln (4\gamma),\\
    e^{2h} &= 4 \gamma {a_{,\zma}}^2 e^{2a} 
    = \gamma \frac{{\epsilon_{,\zma}}^2}{\epsilon} \left.e^{2a}\right|_{\rm cyl.} 
\end{align}
for constant $\gamma$ using $\left({e^{2a}}\right)_{,\zma} = \left.e^{2a}\right|_{\rm cyl.} \epsilon_{,\zma} = 2 a_{,\zma} e^{2a} = 2a_{,\zma} \left.e^{2a}\right|_{\rm cyl.} \epsilon$.
This results in the $\zma$-dependent line element
\begin{align}
    \diff s^2 &= - \epsilon \left(C\ln \frac{\rma}{R}\right) c^2 \diff t^2 + \frac{\beta}{\rma^2\sqrt{C\ln \frac{\rma}{R}}} \left(\diff \rma^2 + \rma^2 \diff \varphi^2\right) + \gamma {\epsilon_{,\zma}}^2 \epsilon^{-1}\left(C\ln \frac{\rma}{R}\right) \diff \zma^2\label{eq:result-withz}
\end{align}
for a $\zma$-dependent function $\epsilon$ and constants $\beta, \gamma, C, R$.

\section{Comparison to Observations}\label{app:obs}

The constants in the cylinder line element, \cref{eq:result-withoutz-R}, remain to be fixed by comparison to observations. To this end, we need 
to consider a situation of cylindrically distributed matter, as given in the case of the visible matter in a galaxy. Note that the derivation of the line element and its low-velocity limit does not make any assumption on that and applies both with and without dark matter given the appropriate symmetries. We will now consider the rotational velocities on galaxies and compare them to the results from the apporximate cylinder line element in the low-velocity limit, as derived in \cref{sec:particles}. This can also be compared to other approaches in terms of the effective acceleration, as shown in \cref{fig:compare-models}.  

\subsection{Baryonic Tully-Fisher Relation}\label{app:btfr}

Another relation that applies in the context of the cylindrically distributed visible matter in galaxies and that can thus be used to constrain the constants is the baryonic Tully-Fisher relation \citep[bTFR,][]{TullyFisher1977A&A....54..661T}. 
This empirical relation states that 
the asymptotic velocity $V_\infty$, i.e. the rotational velocity approached for $\rma \to \infty$ by the observed flat rotation curve of a disk galaxy, and its baryonic mass $M$ are related by the power-law
\begin{align}
    V_\infty^\kappa &\propto M\\
    \left(\frac{V_\infty}{c}\right)^\kappa &= \frac{\mu}{4} GM\label{eq:theor-btfr}
\end{align}
for a phenomenologically defined proportionality constant $\mu$ and exponent $\kappa \approx 4$. 
The relation is commonly assumed to be sustained by the dark matter halo. However, as we do not need dark matter to explain flat rotation curves in the cylinder spacetimes, we consider the bTFR due to the baryonic matter only in these spacetimes and without any dark matter. 
In the regime far from the centre of mass but within the disk of the galaxy, the cylinder solution applies. Thus, we find 
\begin{align}
    V_\infty = \frac{c}{\sqrt{2}}\sqrt[4]{\mu GM}.
\end{align}
to be compared to \cref{eq:flat} which was derived from the approximate effective acceleration obtained from the line element \cref{eq:result-approx}, we can rewrite the constant $C$ in terms of the the proportionality constant $\mu$ or equivalently $\Tilde{\mu} = \mu c^4$:
\begin{align}
    C = \frac{2}{c^2} V_\infty^2 &= \sqrt{\mu G M} = \sqrt{\frac{\Tilde{\mu}}{c^2}\frac{GM}{c^2}}.\label{eq:fix-c-btfr}
\end{align}
The left-hand side of \cref{eq:theor-btfr} is dimensionless and $GM$ is of the dimension of $[L^3 T^{-2}]$, thus the dimension of $\mu$ is $[L^{-3} T^2]$ and the constant $C$ is dimensionless. The alternative formulation in terms of $\Tilde{\mu}$ is easier to interpret as $\Tilde{\mu}$ is of the dimension of an acceleration and the length scale $\frac{GM}{c^2} = \frac{R_{\mathrm{S}}}{2}$ relates to the Schwarzschild radius $R_{\mathrm{S}}$. This acceleration scale can then be discussed in comparison to other models explaining flat rotation curves.


\subsection{Fitting Galaxy Rotation Curves}

\begin{figure}
    \centering
        \includegraphics[width=0.63\textwidth]{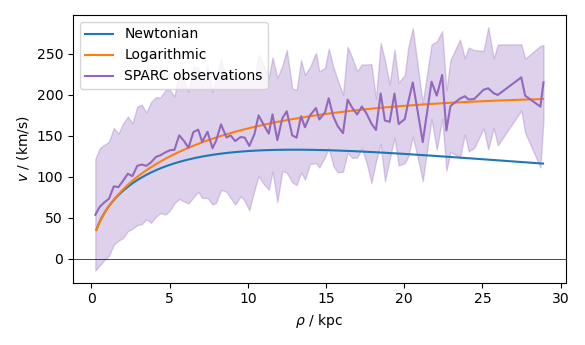}
    \caption{Combined rotation curves from the SPARC catalog \citep{SPARC:2016AJ....152..157L} with corresponding fits. It is conducted based on the exponential density profile, \cref{eq:v-m-enc}, and yields $\rma_0 = (7.2 \pm 0.9) \ \mathrm{kpc}$ and $M_\mathrm{tot} = (10.0 \pm 1.0) \cdot 10^{10} M_\odot$.}
    \label{fig:density-rc}
\end{figure}

The approximate effective acceleration derived in \cref{eq:eff-approx-accel} yields a flat rotational velocity and is thus able to explain flat rotation curves in galaxies with cylindrical symmetry without the need for dark matter. Note however that this derivation is also valid with a dark matter halo, if it fulfills the cylindrical symmetry conditions. 

For a quantitative comparison to observations, the constant $C$ has to be determined, e.g. by use of the bTFR.

Using the asymptotic velocities from the SPARC catalog \citep{SPARC:2016AJ....152..157L}, we can fit the baryonic Tully-Fisher relation (\cref{fig:btfr}) to obtain the constants in the potential and thus in the line element. A fit based on their mass models \citet{SPARC:2016AJ....152..157L}, results in
\begin{align}
    \log_{10}\left(\frac{M}{M_\odot}\right) = (3.64 \pm 0.07) \cdot \log_{10}\left(\frac{V_\infty}{{\rm km / s}}\right) + (2.43 \pm 0.15)
\end{align}
This agrees within 1-2$\sigma$ with the fits in their paper and the resulting the exponent in the power-law, which is within $1\sigma$ of the assumption $\kappa = 4$. Expressing the coefficient of the power-law in terms of the acceleration scale $\Tilde{\mu}$ as given in \cref{eq:fix-c-btfr}, we find
\begin{align}
    \Tilde{\mu} = (1.1 \pm 0.4) \cdot 10^{-10} \frac{\mathrm{m}}{\mathrm{s}^2}\label{eq:value-btfr}.
\end{align}
Notably, this value agrees with the MOND acceleration scale \citep{McGaugh2016PhRvL.117t1101M}, although not by construction. Deriving the bTFR from the MOND interpolation function (\cref{app:bTFR-MOND}) yields an additional factor of $4$, however, the MOND acceleration scale given by \citet{McGaugh2016PhRvL.117t1101M} has a large systematic error allowing for this.

All of these considerations have been derived with a total mass located at the centre. Changing this assumptions for an exponential density profile with scale $\rma_0$ and thus the enclosed mass distribution given by
\begin{align}
    M_{\mathrm{enc}}(\rma) &= M_\mathrm{tot} \left[1 - \frac{\rma + \rma_0}{\rma_0}e^{-\frac{\rma}{\rma_0}} \right],\label{eq:m-enc}
\end{align}
we find 
\begin{align}
    v(\rma) &= \left(\frac{\mu c^4}{4}  G M_\mathrm{tot} \left[1 - \frac{\rma + \rma_0}{\rma_0}e^{-\frac{\rma}{\rma_0}} \right]\right)^\frac{1}{4}.\label{eq:v-m-enc}
\end{align}
This approaches the flat rotation curve for $\rma$ values much greater than the $\rma_0$ scale. The correction is only relevant near the centre and $\rma_0$ gives a radial scale for a bulge (right panel of \cref{fig:density-rc}).

Furthermore, at very large radii, the exponential cutoff in \cref{eq:eff-cutoff-accel} has to be considered and causes the flat rotation curve to decay. However, this behaviour has not been observed so far \citep{mistele2024ApJ...969L...3M}, suggesting the length scales $\frac{1}{\lambda}$ and $\frac{1}{\nu}$ to be very large.

\subsection{Comparison to the MOND Acceleration Scale}
\label{app:bTFR-MOND}

\begin{figure}
    \centering
    \includegraphics[width=0.48\textwidth]{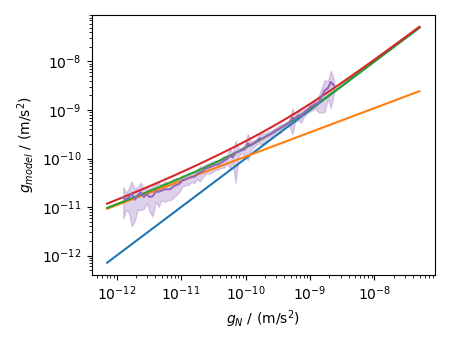}
    \includegraphics[width=0.48\textwidth]{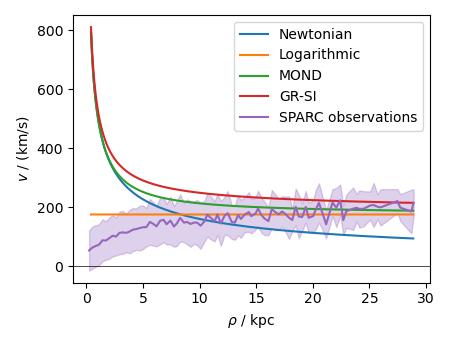}
    \caption{Functional form of the effective acceleration $g_{\rm model} = |\nabla \phi|$ for arbitrary coefficients plotted as a function of the Newtonian acceleration, $g_N$, at the respective radius (left) and the corresponding rotation curves, $v = \sqrt{g \rma}$, according to \cref{eq:flat} with $\rma$ the radial coordinate in the $\zma = 0$ plane (right). The models considered in the plot are the Newtonian potential, the logarithmic potential following from the approximate cylinder solution, the MOND model by McGaugh et al. \citep{McGaugh2016PhRvL.117t1101M} and the GR-SI model by Deur et al. \citep{deur_2020}. For comparison, the mean and standard deviation for rotation curves from the SPARC catalog \citep{SPARC:2016AJ....152..157L} are shown. Note that the logarithmic potential from the cylinder solution is only applicable beyond the horizon $R$, thus it is not expected to match observations for low $\rma < R$ and high $g_N > \frac{GM}{R^2}$ values.}
    \label{fig:compare-models}
\end{figure}

\citet{McGaugh2016PhRvL.117t1101M} found the functional form 
\begin{align}
    g_{\rm MOND}(g_{\rm bar}) &= \frac{g_{N}}{1 - e^{-\sqrt{\frac{g_{N}}{g_\dagger}}}}\\
    g_{\rm MOND}(z) &= g_\dagger \frac{z}{1 - e^{-\sqrt{z}}}
\end{align}
to describe the relation of the observed ($g_{\rm model}$ in \cref{fig:compare-models}, $g_{\rm MOND}$ in this case) and the expected baryonic (Newtonian, $g_N$) accelerations based on the MOND acceleration scale
\begin{align}
    g_\dagger = (1.20 \pm 0.02 \text{ (stat.)} \pm 0.24 \text{ (syst.)}) \cdot 10^{-10} \frac{\rm m}{\rm s^2}\label{eq:MOND-scale}
\end{align}
\citep{McGaugh2016PhRvL.117t1101M} and $z = \frac{g_{N}}{g_\dagger}$. In the non-Newtonian limit, $z \ll 1$, this approaches
\begin{align}
    g_{\rm MOND} \approx g_\dagger \sqrt{z} = \sqrt{g_\dagger g_{N}}
\end{align}
which corresponds to a rotational velocity of 
\begin{align}
    \frac{m v^2}{r} &= m \sqrt{g_\dagger g_{N}}\\
    v^4 &= r^2 g_\dagger \frac{GM}{r^2}\\
    v^4 &= g_\dagger GM\label{eq:mond-btfr}
\end{align}
using $g_{N} = \frac{GM}{r^2}$. In contrast to this, the baryonic Tully-Fisher relation in its form given in \cref{eq:theor-btfr} yields
\begin{align}
    v^4 &= \frac{\Tilde{\mu}}{4} GM
\end{align}
where $\Tilde{\mu}$ is phenomenologically found to agree with $g_\dagger$ in \cref{eq:value-btfr}. The rotational velocity in \cref{eq:mond-btfr} found from the limiting behaviour of the MOND interpolation function thus differs from the observations of the bTFR by a factor of $4$. The large systematic error in the MOND acceleration scale, \cref{eq:MOND-scale}, allows for this.

\section{Newtonian Gravity for Line-Like Sources}\label{app:newt}

For an extended line-like source in Newtonian gravity, i.e. with the density $\varrho$ given by 
\begin{align}
    \varrho(\ran, \zan) \propto \frac{\delta(\ran)}{\ran}\Theta(a + \zan)\Theta(a - \zan),
\end{align}
we find
\begin{align}
    \phi(\ran, \zan) &= \int \Tilde{\ran} \ \diff \Tilde{\ran} \ \diff \varphi \ \diff \Tilde{\zan} \ \frac{\varrho(\Tilde{\ran})}{\sqrt{(\Tilde{\ran} - \ran)^2 + (\Tilde{\zan} - \zan)^2}}\\
    &\propto 2\pi \int_{-a}^a \diff \Tilde{\zan} \frac{1}{\sqrt{\ran^2 + (\Tilde{\zan} - \zan)^2}}
\end{align}
which yields
\begin{align}
    \phi(\ran, 0) &\propto 
    \frac{1}{2} \left.\left(-\ln\left(1 - \frac{\zan'}{\sqrt{\ran^2 + \zan'^2}}\right) + \ln\left(1 + \frac{\zan'}{\sqrt{\ran^2 + \zan'^2}}\right)\right)\right|_{-a}^{a}\\
    &= -\ln\left(1 - \frac{a}{\sqrt{\ran^2 + a^2}}\right) + \ln\left(1 + \frac{a}{\sqrt{\ran^2 + a^2}}\right)\\
    &= \ln \left(\frac{\ran^2 + a^2 + 2a\sqrt{\ran^2 + a^2} + a^2}{\ran^2 + a^2 - a^2}\right)\\
    &= \ln \left(1 + 2\frac{a}{\ran}\left(\frac{a}{\ran} + \sqrt{1 + \left(\frac{a}{\ran}\right)^2}\right)\right)
\end{align}
within the disk. Two limits are important when investigating this potential. First, consider $\frac{a}{\ran} \gg 1$, i.e., close to the centre of mass at $\ran = 0$, and obtain
\begin{align}
    \phi(\ran, 0) &\appropto \ln \left(4\frac{a^2}{\ran^2}\right)\\
    &= -2 \ln\left(\frac{\ran}{2a}\right)
\end{align}
which is of logarithmic form. 
Conversely, the limit far from the centre, i.e. $\frac{a}{\ran} \ll 1$, yields
\begin{align}
    \phi(\ran, 0) &\appropto \frac{2a}{\ran}\left(\frac{a}{\ran} + \sqrt{1 + \left(\frac{a}{\ran}\right)^2}\right) + \mathcal{O}\left(\left(\frac{a}{\ran}\right)^2\right)\\
    &= \frac{2a}{\ran} + \mathcal{O}\left(\left(\frac{a}{\ran}\right)^2\right)
\end{align}
which reduces to the Newtonian potential.

\end{document}